\documentclass[aps,pra,twocolumn,footinbib, superscriptaddress]{revtex4-2}

\usepackage{graphicx}
\usepackage{hyperref}
\usepackage{amsmath,amssymb, mathtools, physics, dsfont}
\usepackage{orcidlink}
\usepackage{microtype}
\usepackage[T1]{fontenc}

\usepackage[dvipsnames]{xcolor}

\usepackage[markup=underlined]{changes}

\definechangesauthor[name=Miki,color=BrickRed]{ed}

\hypersetup{
	colorlinks=true,       
	linkcolor=Brown,        
	citecolor=Blue,        
	filecolor=magenta,     
	urlcolor=RoyalPurple,      
}

\begin{document}

\title{Quantum Gates via Dynamical Decoupling of Central Qubit \\ on IBMQ and \textsuperscript{15}NV Center in Diamond}

\author{Lucas Tsunaki \orcidlink{0009-0003-3534-6300}}
\affiliation{Department Spins in Energy Conversion and Quantum Information Science (ASPIN), Helmholtz-Zentrum Berlin für Materialien und Energie GmbH, Hahn-Meitner-Platz 1, 14109 Berlin, Germany}

\author{Michael Dotan \orcidlink{0009-0005-8690-3394}}
\affiliation{Department Spins in Energy Conversion and Quantum Information Science (ASPIN), Helmholtz-Zentrum Berlin für Materialien und Energie GmbH, Hahn-Meitner-Platz 1, 14109 Berlin, Germany}

\author{Kseniia Volkova \orcidlink{0009-0000-7258-8921}}
\affiliation{Department Spins in Energy Conversion and Quantum Information Science (ASPIN), Helmholtz-Zentrum Berlin für Materialien und Energie GmbH, Hahn-Meitner-Platz 1, 14109 Berlin, Germany}

\author{Boris Naydenov \orcidlink{0000-0002-5215-3880}}
\email{boris.naydenov@helmholtz-berlin.de}
\affiliation{Department Spins in Energy Conversion and Quantum Information Science (ASPIN), Helmholtz-Zentrum Berlin für Materialien und Energie GmbH, Hahn-Meitner-Platz 1, 14109 Berlin, Germany}
\affiliation{Department of Physics, Freie Universität Berlin, Arnimallee 14, 14195 Berlin, Germany}

\date{\today}

\begin{abstract}

\noindent We demonstrate a hardware-agnostic protocol for realizing fast, high-fidelity gates through dynamical decoupling (DD) pulse sequences applied to a central qubit coupled to target qubits.
The target qubits are controlled by leveraging their intrinsic interaction with the central qubit, eliminating the need for slow, error-prone direct control.
We develop and implement the DD-gate protocol within two distinct frameworks: a general model with minimal assumptions, benchmarked on a gate-based digital quantum simulator given by the IBMQ; and an experimentally realistic case with a nitrogen-15 vacancy center ($^{15}$NV) in diamond.
Using IBMQ, we are able to elucidate the underlying quantum dynamics of the DD-gates and test them, independently of experimental constraints.
For $^{15}$NV, we realize the protocol considering system-specific properties, which could represent a significant reduction in gate duration and improved technological scalability compared with current dynamical-decoupling-based control.
We also propose a simple application for high-efficiency polarization of the $^{15}$N nuclear spin that could potentially be less technically demanding than current methods.
Altogether, this work provides a robust strategy for quantum control that can be implemented in arbitrary systems fitting the central-target qubit architecture.
Beyond these results, our open-source simulations and implementations for both platforms provide a practical framework for simulating time-dependent qubit dynamics on NISQ-era gate-based quantum processors.

\end{abstract}

\maketitle


\section{Introduction}

\begin{figure*}[t!]
	\includegraphics[width=\textwidth]{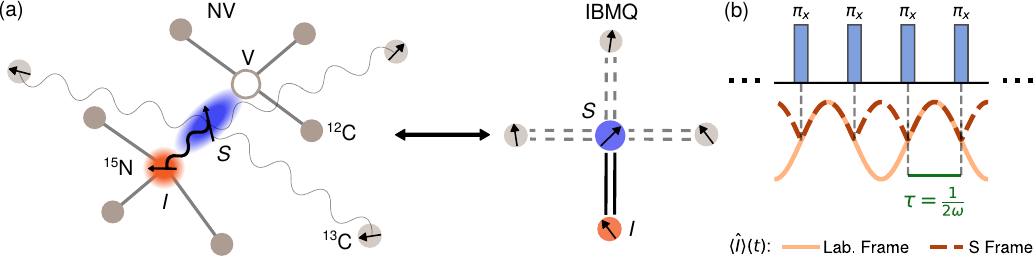}
	\includegraphics[width=\textwidth]{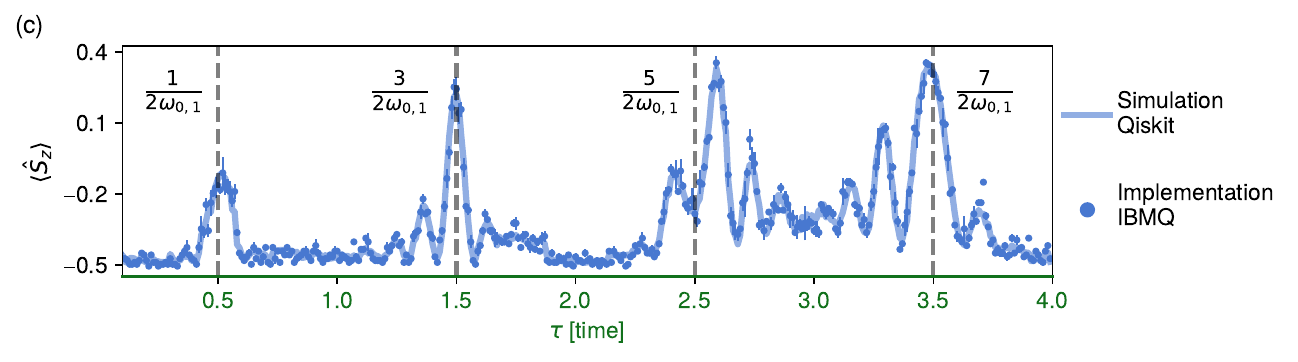}
	\caption[IBMQ Quantum Simulator for DD-gates with NVs]{\textbf{(a)} The $^{15}$NV center in the diamond lattice consists of an electron spin $S$ (control qubit) and the $^{15}$N nuclear spin $I$ (target qubit). These are surrounded by a $^{13}$C nuclear spin bath.
	The IBMQ is employed as a digital quantum simulator to emulate these dynamics, without complex bath interactions and other experimental constraints.
	\textbf{(b)} CPMG dynamical decoupling sequence.
	By changing the separation $\tau$ between consecutive $\pi_x$ pulses applied on the control qubit, the sequence filters out signals which are not in resonance with the frequency $\omega=1/(2\tau)$ and its odd multiples.
	In this way, the target-qubit expectation value $\langle \hat{I} \rangle(t)$ averages to zero in the laboratory frame, but adds constructively in the control qubit's reference frame.
	\textbf{(c)} Demonstrated and simulated time evolution of the control qubit under a CPMG-10 sequence in IBMQ.
	The $\langle \hat{S}_z \rangle$ observable shows resonances due to its interaction with the target qubit at odd multiples of $1/(2\omega_{0, 1})$, with pronounced sidebands at larger $\tau$ values.
	}
	\label{fig:intro}
\end{figure*}

The generation of fast, reliable, and technologically scalable multi-qubit gates is a fundamental challenge for quantum computing~\cite{CROT_NV2} and other quantum technology applications~\cite{sensing_NV1, sensing_NV2, sensing_NV3, communication_NV1, memory_NV}.
A critical challenge in the implementation is the inherent trade-off between gate fidelity and operation time~\cite{15N_gates}.
Here, we demonstrate a method that leverages dynamical decoupling (DD)~\cite{DD_NV} to mediate multi-qubit gates, simultaneously optimizing for speed and fidelity, while preserving technological scalability.

Our approach is designed for, but not restricted to, a common architecture within quantum technology platforms: a central `control' qubit that strongly interacts with external fields, coupled to one or more `target' qubits with weak direct control.
Here, coherent control of the central qubit is usually much faster and more precise than that of the coupled qubits.
In contrast, the weakly interacting qubits generally have much longer coherence than the central qubit, placing them as ideal candidates for quantum memories~\cite{memory_NV, qtoken_1}.
A primary example of such systems is color center defects in solids~\cite{CC1, CC2}, which commonly have an electron spin ($S$) coupled to one or more nuclear spins ($I$), where gate durations for the former are several orders of magnitude shorter than for the latter~\cite{CROT_NV1}.
Although direct excitation of the weakly interacting qubits by resonant pulses is possible~\cite{CROT_NV1}, an alternative that makes use of fast, high-fidelity gates on the central qubit is highly desirable.
Within this context, DD techniques exploit the intrinsic interaction between the control and target qubits to realize gates.

Pulsed DD sequences consist of periodic reversals of the quantum evolution of a qubit, so that environmental noise is canceled during the total time evolution.
In particular, the CPMG-$N$ sequence~\cite{CPMG} consists of $N$ repeated $\pi_x$ pulses separated by free-evolution intervals of duration $\tau$.
A $\pi_y/2$ pulse is also applied before the sequence to drive the qubit from its initial state $\ket{0}$ at the pole of the Bloch sphere to the equator, where the interaction with the target qubits can be probed.
Similarly, a $\pi_y/2$ pulse is applied at the end of the sequence to project the acquired phase back to its quantization axis $z$, where it can be measured. 
As the number of $\pi$ pulses $N$ increases, the sequence acts as a narrowband filter for frequencies corresponding to twice the inverse pulse separation $\omega = 1/(2\tau)$ and odd multiples of it~\cite{multipulse1, spurious}, see Fig.~\ref{fig:intro}~(b).
In the static laboratory frame, the target qubits perform Larmor precessions characterized by periodic oscillations of their expectation values $\langle \hat{I} \rangle (t)$, which average out to zero.
In contrast, in the reference frame of the control qubit, the expectation values $\langle \hat{I} \rangle (t)$ add up constructively if the pulse separation is resonant with its Larmor frequency, while other signals with different frequencies cancel out destructively.
Thus, by tuning the pulse separation to the resonant frequency of one of the target qubits, all other interactions in the system can be filtered from the control qubit, apart from the interaction with the target, which we then use to perform gates on the target qubit without applying slow resonant pulses to it.

This application is notably well suited for color centers, due to the natural presence of both electron and nuclear spin species.
Specifically, we consider the nitrogen-vacancy (NV) center in diamond~\cite{NV_review1, NV_review2}, composed of a substitutional nitrogen adjacent to a vacancy in the diamond lattice, as shown in Fig.~\ref{fig:intro}~(a).
NVs host an electron spin $S=1$ and a nitrogen nuclear spin $I=1$ for the $^{14}$N isotope (99.6\% natural abundance) or $I=1/2$ for $^{15}$N (0.4\%).
In this study, we only consider the $^{15}$NV center due to the absence of a quadrupole interaction, which suppresses the Larmor precession of the nuclear spin~\cite{quadrupole} (see Sec.~\ref{sec:H_NV}).
From now on, the $^{15}$NV center is simply referred to as NV. 
Apart from the nitrogen, the electron spin of the NV can also couple with spin-1/2 nuclei from $^{13}$C atoms in the diamond lattice~\cite{Nizovtsev10} and other paramagnetic impurities~\cite{P1}.

DD techniques have already been extensively applied to NV centers for a wide range of applications, initially for extending the coherence of the NV's electron spin~\cite{DD_NV} up to nearly a second~\cite{NV_DD_2}.
Another common application is for quantum sensing, where it is used to probe the interaction with the spin bath itself.
Notably, the technique has been utilized to measure weak signals from $^{1}$H nuclei at the diamond surface~\cite{DD_H_1, DD_H_2} and from $^{13}$C isotopes in the lattice~\cite{CROT_NV_DD, CROT_13C_2}, where background signal was present.
DD with NVs has also been combined with resonant radio-frequency (RF) pulses to the $^{13}$C nuclear spins to generate multi-qubit gates in the NV-$^{13}$C system~\cite{CROT_13C_3, CROT_13C_4, CROT_13C_5, CROT_13C_6}, being of great interest for analog quantum-simulator proposals~\cite{qsimulator} and quantum memories~\cite{CROT_13C_1}.
However, the use of DD sequences for control of the native $^{15}$N nuclear spin of the NV without direct resonant RF pulses remains an open research problem.
Here, we demonstrate the realization with the $^{15}$N spin, without adoption of rotating frames~\cite{beyond_RWA, slichter} and using only microwave (MW) pulses on the electron spin.

Using the $^{15}$N nuclear spin instead of $^{13}$C has a significant technological advantage with regard to scalability, as the interaction Hamiltonian for a coupled $^{15}$N is the same for all NVs.
$^{13}$C nuclei are stochastically distributed in the diamond lattice around the NV at current technological levels of material fabrication, resulting in a variety of spin Hamiltonians~\cite{13C_fam}.
The scalability of such a method is particularly useful for diamond-based quantum token proposals~\cite{qtoken_1, qtoken_2}, where the nuclear spins are envisioned to store quantum states prepared using the electron spin as an auxiliary qubit.
In addition, our theoretical model goes beyond previous implementations, by allowing generalized coupling strengths that would not be suitable in perturbative approaches with adoptions of rotating frames, such as the rotating wave approximation (RWA).

Although particularly interesting for color centers, the use of DD of a control qubit for performing quantum gates to target qubits could also be relevant to other  architectures, as superconducting circuits or quantum dots.
It is thus indispensable to develop a hardware-agnostic non-perturbative model for verifying the fundamental principles of the technique, while benchmarking it on a general-purpose quantum processor.
With this goal, we employ IBM Quantum Platforms (IBMQ) composed of superconducting qubits~\cite{hardware1, hardware2, hardware3} as digital quantum simulators.
The IBMQ can thus emulate the behavior of an NV center [Fig.~\ref{fig:intro}~(a)], or other general quantum systems, free from complex spin bath dynamics and other experimental constraints.

We begin this work in Sec.~\ref{sec:IBMQ} by introducing a minimal Hamiltonian model for DD-mediated gates with generalized quantum systems and by implementing it on IBMQ using Qiskit~\cite{qiskit}.
This proof of concept demonstrates the feasibility of the DD-filtered drive on a digital quantum processor.
Building on this general model, we then discuss its application to NV centers in Sec.~\ref{sec:NV}, considering a more complex Hamiltonian model than for IBMQ.
Experimental measurements of the NV's electron spin are shown and compared with simulations performed using QuaCCAToo software~\cite{quaccatoo_paper, quaccatoo_git}.
An application of this technique for $^{15}$N nuclear spin polarization is proposed, which could significantly simplify experimental complexity compared with existing methods~\cite{single_shot_readout, 15N_gates, DNP1, DNP2}.
Finally, we conclude in Sec.~\ref{sec:conclusion} by discussing the limitations and possible improvements of the model.
The simulations and implementations with IBMQ are open source, provided through a Python package available at~\cite{dd_gates_git}.
In addition, the simulations of the NV center are provided in the QuaCCAToo repository~\cite{quaccatoo_git}.

\section{Universal Model Benchmarked on IBMQ}\label{sec:IBMQ}

To realize multi-qubit gates via DD of the central qubit, we make use of the IBMQ Torino processor from the Heron family (see Appendix~\ref{appendix:qiskit} for more details).
In Sec.~\ref{sec:IBMQ_H0}, we introduce the theoretical framework for performing gate-based quantum simulation of time-dependent qubit dynamics in the laboratory frame.
This framework has potential applications beyond DD-gates for general quantum simulation on noisy intermediate-scale quantum (NISQ) hardware.
Building on that, in Sec.~\ref{sec:IBMQ_2} we demonstrate the DD-gate with the central qubit coupled to one target qubit.
Finally, in Sec.~\ref{sec:IBMQ_3}, we realize the gate with the central qubit coupled to two target qubits, demonstrating the robustness of the DD method.
In addition, Appendix~\ref{appendix:Azx} discusses the effect of the coupling factor $A_{zx}$ and Appendix~\ref{appendix:pulse_errors} analyzes the effects of pulse errors in the DD sequences.


\subsection{Theoretical Framework}\label{sec:IBMQ_H0}

A general system for DD-gates comprises the central qubit, represented by the spin operators $\hat{\mathbf{S}}=(\hat{S}_x, \hat{S}_y, \hat{S}_z)$, and $M$ target qubits, given by the operators $\hat{\mathbf{I}}_j=(\hat{I}_{x,j}, \hat{I}_{y,j}, \hat{I}_{z,j})$ with indices $j=1,\; 2, \;..., \;M$.
We take all qubits initialized in the $\ket{0}$ state and having their quantization axes along $z$, with the central qubit having a Larmor frequency $\omega_{0,0}$ that gives the energy difference between its states.
The target qubits have Larmor frequencies $\omega_{0,j}$, which are much smaller than that of the central qubit, $\omega_{0,j} \ll \omega_{0,0}$.
We further consider that each target qubit has a coupling with the central qubit described by a tensor $\mathbf{A}^j$, where interactions among target qubits are neglected.

Using these considerations, the system's Hamiltonian can be written as
\begin{equation}\label{eq:IBMQ_H0_A}
	\hat{H}_0 = \omega_{0, 0}\hat{S}_{z}
	+ \sum_{j=1}^{M} \left( \omega_{0, j} \hat{I}_{z, j}
	+ \hat{\mathbf{S}} \cdot \mathbf{A}^j \cdot \hat{\mathbf{I}}_j \right) ,
\end{equation}
where we assume $h=1$ throughout this work and take the Hamiltonians and coupling terms in arbitrary units of [frequency], corresponding to the inverse of [time].
For the generation of the DD-gates, we are only interested in the coupling terms that involve the $\hat{S}_z$ and $\hat{I}_{x,j}$ components in the form of $A_{zx}^j\hat{S}_z\hat{I}_{x,j}$.
This is because the central qubit can sense interactions along its quantization axis $z$~\cite{ambiguous_resonances} when it is driven to the equator of the Bloch sphere, while the perpendicular components $x$ and $y$ cause qubit flips.
Contrarily, in the target qubits, we are interested in these perpendicular components that will cause transitions between the states, where we only take the $x$ component due to the freedom of choice of the coordinate system.
Therefore, in the minimal Hamiltonian model, we have the coupling interactions being given by $\hat{\mathbf{S}} \cdot \mathbf{A}^j \cdot \hat{\mathbf{I}}_j \rightarrow A_{zx}^j \hat{S}_{z} \hat{I}_{x, j}$.
Strictly speaking, the interaction tensor $\mathbf{A}^j$ should be symmetrical and we would need to have terms with $A_{zx}^j \hat{S}_{x} \hat{I}_{z, j}$.
However, these terms do not significantly affect the system dynamics of the system under the DD sequences~\cite{CROT_NV_DD} and increase the computational costs with IBMQ, thus being neglected.
NV's Hamiltonian is more complex, but it still shows these same features, as discussed in detail in Sec.~\ref{sec:H_NV} and Appendix~\ref{appendix:NV_PAS}.

Now we model the interaction of the central qubit with the control field. Considering harmonic square pulses along the $x$ direction~\cite{quaccatoo_paper}, the time-dependent Hamiltonian in the laboratory frame during a pulse is simply given by
\begin{equation}\label{eq:IBMQ_H1}
	\hat{H}_1(t) = \omega_{1} \cos \left( 2 \pi \omega_p t + \phi \right)\hat{S}_{x} ,
\end{equation}
where $\omega_1$ is the Rabi frequency of the transition~\cite{rabi}, $\omega_p$ is the frequency of the pulse in resonance with $\omega_{0, 0}$ and $\phi$ its phase, which controls the rotation axis in the rotating frame.
By changing the pulse duration $t_p$, the central qubit performs Rabi oscillations~\cite{rabi_original} of the $\langle \hat{S}_z \rangle $ observable between its two eigenstates, where for $t_\pi = 1/(2\omega_1)$, we have complete inversion of the populations, i.e. a $\pi$-pulse.

For the description of single qubit dynamics, a rotating frame is typically adopted~\cite{slichter}, where $\hat{H}_1(t)$ becomes time-independent.
However, the unitary transformation operator of the rotating frame does not necessarily commute with $\hat{H}_0$ from Eq.~\ref{eq:IBMQ_H0_A} depending on $\mathbf{A}^j$, which would then lead to time dependencies in the rotating frame (as is the case with the NV Hamiltonian).
Furthermore, the use of the rotating frame imposes that $ \omega_{0, 0} \gg A_{zx}^j$ and $ \omega_{0, 0} \gg \omega_{1}$, which are not always valid, or in the second case advantageous due to a slow driving of the central qubit.
Altogether, the most general approach for this problem is to describe the dynamics of the quantum simulator in the static laboratory frame, in contrast to most quantum computing applications.
This time-dependent description of the control Hamiltonian $\hat{H}_1(t)$ also allows us to directly account for pulse errors (Appendix~\ref{appendix:pulse_errors}). 

Implementing a free-evolution under $\hat{H}_0$ in the IBMQ processor at the laboratory frame is a straightforward operation.
For that, the state of the system can be evolved with the time-evolution operator
\begin{equation}\label{eq:Utau}
	\hat{U}_\tau = \exp \left[- i 2 \pi ( \tau - t_\pi)  \hat{H}_0  \right] ,
\end{equation}
where $\tau - t_\pi$ is the actual pulse separation when considering realistic pulses with finite length~\cite{quaccatoo_paper}.
The exponential operators are applied to the qubits through Pauli evolution gates~\cite{pauli_evolution}, transpiled to the native gates of the hardware.
More implementation details on the transpilation and execution of the circuits are given in Appendix~\ref{appendix:qiskit}, while the exact implementation of the time-evolution gates can be found in~\cite{dd_gates_git}.

If we wanted to analytically solve the system dynamics during the pulse applications under the time-dependent Hamiltonian $\hat{H}_0 + \hat{H}_1(t)$, a Dyson series~\cite{dyson_series} could be used, involving an infinite sum of time integrals.
But clearly, when considering this operation with a digital quantum simulator, the time variable cannot be continuously divided into infinitesimal components $dt$, infinitely integrated and summed.
Instead, to achieve the pulse dynamics with IBMQ, we perform a Trotterization decomposition~\cite{trotterization1, trotterization2} of the continuous time-evolution into a finite number of gates with $\Delta t$ steps as
\begin{equation}\label{eq:Up}
	\hat{U}_p(t_p) \approx \prod_{k=1}^{t_p/\Delta t}
	\exp \left\{-i 2 \pi \Delta t\left[\hat{H}_0 + \hat{H}_1(t_k)\right] \right\},
\end{equation}
ordered from right to left.
For this approximation to be valid, $\Delta t$ needs to be small compared to $1/\omega_{0, 0}$, such that the Hamiltonian is nearly constant at the time $t_k$ and it can be applied as if it was time-independent.
However, even in the continuous case of the Dyson series, the perpendicular coupling terms and the time-dependent drive make a general analytical solution to the time-evolution operator unreachable.

In the end of the sequence, we measure the central and target qubits by their quantum mechanical observables $\hat{S}_z$ and $\hat{I}_{z,j}$.
In our model, non-unitary processes such as decoherence and relaxation are not explicitly accounted, which could be numerically incorporated through collapse operators in the Lindblad equation~\cite{lindblad, quaccatoo_paper}.
Instead, they are included within the noise model of the IBMQ processors in the Qiskit simulations framework (Appendix~\ref{appendix:qiskit}).
As will be seen in the following section, decoherence has a negligible effect in IBMQ at these time-scales, but does play an important role for NVs.


\subsection{One Target Qubit Gate}\label{sec:IBMQ_2}

\begin{figure}[b!]
	\includegraphics[width=\columnwidth]{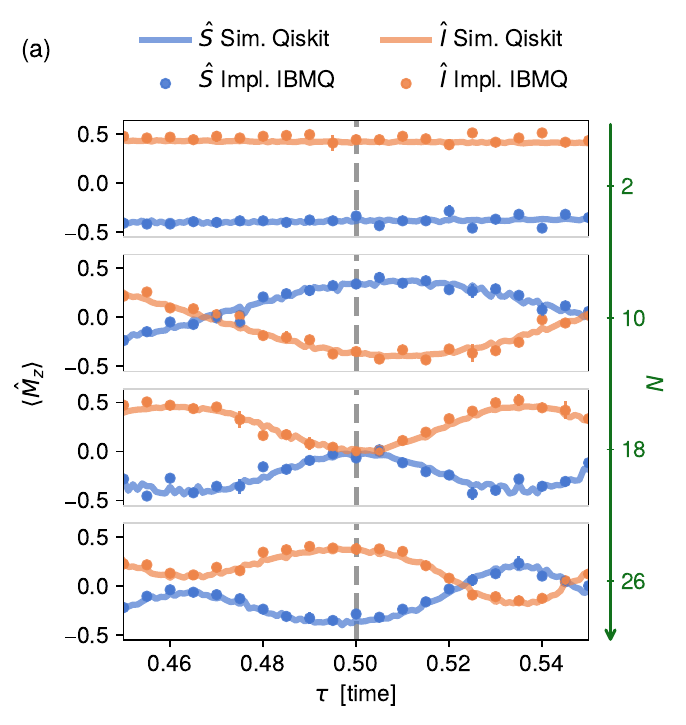}\vspace{-.3cm}
	\begin{flushleft}
		\includegraphics[width=.9\columnwidth]{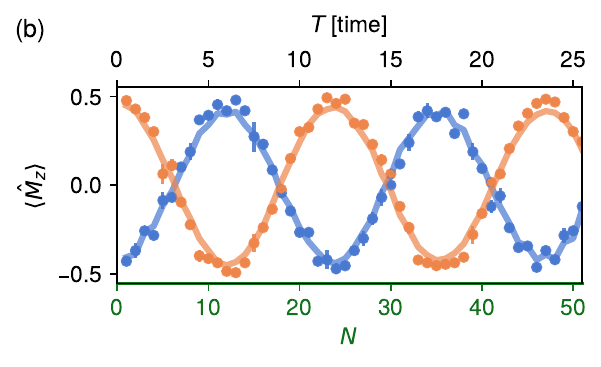}
	\end{flushleft}
	\caption[DD Spectrum Resonances from Target Qubit in IBMQ]{
		\textbf{(a)} CPMG-$N$ sequences applied to the control qubit implemented in IBMQ and simulated with Qiskit.
		The target qubit observable $\langle \hat{I}_z \rangle$ has a time-evolution opposite to the control qubit $\langle \hat{S}_z \rangle$, with a resonance at $\tau=1/(2\omega_{0, 1})=0.5$~[time].
		As the number of pulses $N$ increases, the amplitude of the resonance varies and sidebands become more pronounced.
		\textbf{(b)} Rabi-like oscillations of the central and target qubits.
		By taking  $\tau=0.5$~[time] fixed and varying $N$, the two observables perform opposite harmonic oscillations, due to opposite phase accumulations at the equator of the Bloch sphere by the end of the DD sequence.
		This variation of the $\langle \hat{I}_z \rangle$ observable is the basis for the DD-gates.
	}
	\label{fig:cpmg_ibmq}
\end{figure}

\begin{figure*}[t!]
	\includegraphics[width=.49\textwidth]{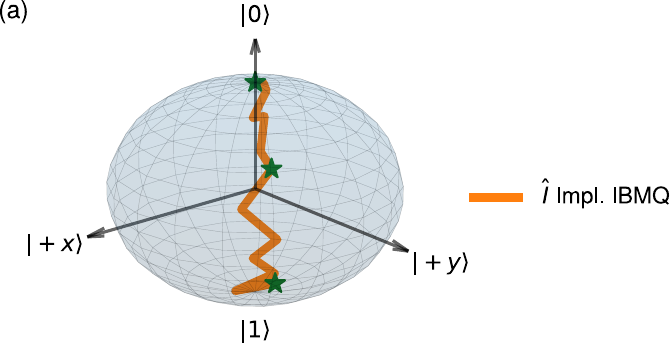}\hfill
	\includegraphics[width=.49\textwidth]{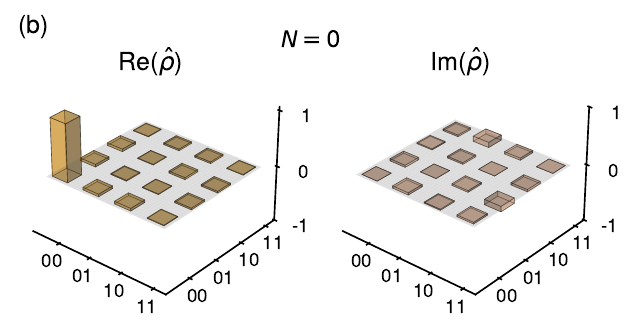}
	\includegraphics[width=.49\textwidth]{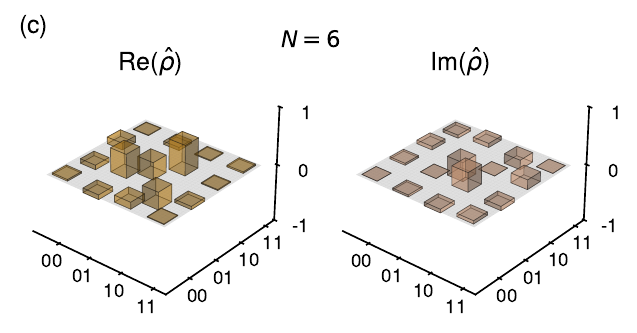}\hfill
	\includegraphics[width=.49\textwidth]{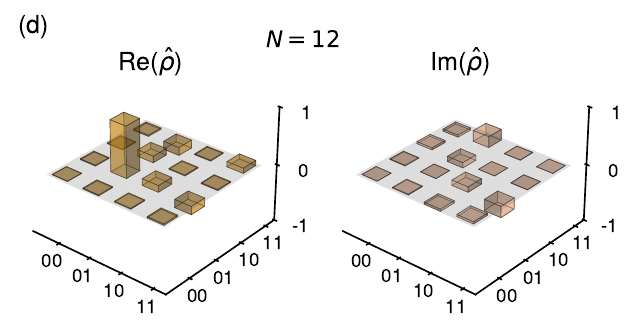}
\caption[Quantum State Tomography of Target Qubit under DD-Gate in IBMQ]{
		\textbf{(a)} Bloch sphere representation of the target qubit under the DD-gate implemented with IBMQ.
		During the CPMG-$N$ sequences, the subsystem state obtained from a partial trace of the two-qubits occupies points inside the sphere, which does not qualify the gate as a proper rotation.
		Full density matrices $\hat{\rho}$ at \textbf{(b)} $N=0$, \textbf{(c)} $N=6$ and \textbf{(d)} $N=12$ pulses, demonstrated by quantum state tomography.
		These points are represented in (a) by the green stars.
		At $N=0$ the system is well initialized in the $\ket{00}$ state, with small readout errors.
		In the middle of the gate at $N=6$, the diagonal population term is distributed into off-diagonal terms representing entanglement, which results in mixed states for the single qubits when taking the partial traces of $\hat{\rho}$.
		At $N=12$, the entanglement terms are reverted back into diagonal terms, resulting in a population inversion for the target qubit with fidelity 0.948, limited by the discretized nature of $N$.
	}
	\label{fig:DM_ibmq}
\end{figure*}

We begin by implementing DD-gates with one single coupled qubit.
In this case, Eq.~\ref{eq:IBMQ_H0_A} becomes
\begin{equation}\label{eq:ibmq_H0_2q}
	\hat{H}_0 = \omega_{0, 0}\hat{S}_{z}
	+ \omega_{0, 1} \hat{I}_{z}
	+ A_{zx} \hat{S}_{z} \hat{I}_{x} ,
\end{equation}
where we simplify the notation by dropping the $j$ index.
We numerically take $\omega_{0, 0}=50$, and $\omega_{0, 1}=1$ in units of $[\textrm{frequency}]$.
In $\hat{H}_1(t)$ (Eq.~\ref{eq:IBMQ_H1}) we take $\omega_{1}=5$~[frequency], such that $\omega_{0,0} \gg \omega_{1}$ is valid and the driving of the central qubit is harmonic~\cite{beyond_RWA}.
Last, in Eq.~\ref{eq:Up}, we adopt a time-step of $\Delta t=0.001$~[time].
If we assume that the central and target qubit are electron and nuclear spins respectively, the Larmor frequency of the former is typically $10^3$ to $10^4$ times larger than the latter.
On the other hand, a large $\omega_{0, 0}$ would require a small time step $\Delta t$, so that $\hat{H}_1(t)$ is nearly constant in each discretized time $t_k$.
Thus, to avoid increased computational costs in the IBMQ, we set $\omega_{0, 0}/\omega_{0, 1}=50$.

A CPMG sequence with $N=10$ pulses was implemented in IBMQ and simulated in Qiskit with $A_{zx}=0.1$~[frequency], as shown in Fig.~\ref{fig:intro}~(c).
By varying the pulse separation $\tau$, the central qubit observable $\langle \hat{S}_z \rangle$ exhibits multiple resonances centered at the odd multiples of the inverse of the fundamental resonant frequency: $1/(2\omega_{0, 1})$, $3/(2\omega_{0, 1})$, $5/(2\omega_{0, 1})$ and $7/(2\omega_{0, 1})$.
These resonances also have sidebands, which become more pronounced at the longer pulse separations due to the increased interaction time between the two qubits and its resulting effect on the filter function of the DD sequence~\cite{phd_muller}.

This method is broadly used for quantum sensing of coupled qubits~\cite{multipulse1, spurious, CROT_NV_DD}, but here our goal is to use the DD sequence resonances to control the time-evolution of the target qubit.
To characterize these dynamics of the target qubit, we take $\tau$ values around the first resonance at $\tau=1/(2\omega_{0, 1})$ and vary the number of pulses $N$, measuring both observables as presented in Fig.~\ref{fig:cpmg_ibmq}~(a).
Here, we take $A_{zx}=0.2$~[frequency] in order to achieve better resonance contrast (Appendix~\ref{appendix:Azx}).
As the number of pulses increases, the filter function of the DD sequence changes~\cite{phd_muller}, resulting in narrower resonance linewidths, with more pronounced sidebands and oscillations of the resonance amplitudes.
Notably, the target qubit has a time-evolution under the DD sequence opposite to the central qubit, when considering this initial state $\ket{0,0}$ (a case with a mixed initial state is discussed in Sec.~\ref{sec:pol_gen}).
This indicates that, at the end of the DD sequence, the target qubit accumulates a phase of opposite sign and same amplitude as the control qubit at the equator of the Bloch sphere, which is then projected into the quantization axis $z$ by the final $\pi_y/2$-pulse.
Thus an increase in the resonance amplitude of the $\langle \hat{S}_z \rangle$ observable at some $\tau$ corresponds to a decrease by the same amount in the $\langle \hat{I}_z \rangle$ observable of the target qubit.

To further study this effect, we take $\tau=0.5$~[time] fixed and measure the expectation value of the observables as a function of $N$ in Fig.~\ref{fig:cpmg_ibmq}~(b).
The central and target qubits observables $\langle \hat{S}_z \rangle$ and $\langle \hat{I}_z \rangle$ perform opposite harmonic oscillations with the same frequency and amplitude, which do not decay when damping processes~\cite{damping_process} are not present. 
The dependency of this oscillation frequency on the coupling constant is discussed in detail in Appendix~\ref{appendix:Azx}.
By taking $N=12$ pulses, we have a complete inversion of both observables, totaling $T_\pi = N \tau = 6$ units of time, which can represent a considerable reduction in the gate duration compared to direct excitation of the target qubit (see Sec.~\ref{sec:NV_2qubit}).
In all sequences, there is a strong agreement between the simulation and the implementation using the digital quantum simulator, which validates the proposed model. 

The oscillations of the observables would suggest that the system is undergoing a Rabi oscillation~\cite{rabi_original} corresponding to rotations in the Bloch sphere for both qubits, but this is not the case.
By measuring the components $\langle \hat{I}_x \rangle$ and $\langle \hat{I}_y \rangle$ perpendicular to the quantization axis, we observe that in fact the qubits evolve from one pole of the Bloch sphere (state $\ket{0}$) to the other (state $\ket{1}$) through points inside the sphere, not on its surface as it is for the Rabi experiment.
Fig.~\ref{fig:DM_ibmq}~(a) depicts this effect, by showing data for $N$ between 0 to 12 of the reduced density matrix corresponding to the subsystem of the target qubit:
\begin{equation}\label{eq:rho_I}
	\hat{\rho}_{I} =  \Tr_{S} (\hat{\rho} ) = \frac{\hat{\mathds{1}}}{2}
	+ \sum_{a=x,y,z} \langle \hat{\mathds{1}} \otimes \hat{I}_{a} \rangle  \hat{\sigma}_a , 
\end{equation}
where we explicitly write the tensor product of the target qubit observables and $\hat{\sigma}_a$ denotes the Pauli matrices.

When considering single qubits, the fact that the state is represented by a point inside the sphere would indicate a non-unitary dynamics, inducing state mixing.
But with two qubits, this is not a valid interpretation, as the Bloch sphere is an incomplete description of the system.
Specifically, the density matrix for a two-qubit system is
\begin{multline*}
	\hat{\rho} = \frac{\hat{\mathds{1}}}{4}
	+ \frac{1}{2} \sum_{a=x,y,z} \left( \langle \hat{S}_{a} \otimes \hat{\mathds{1}} \rangle \hat{\sigma}_a \otimes \hat{\mathds{1}}
	+ \langle \hat{\mathds{1}} \otimes \hat{I}_{a} \rangle  \hat{\mathds{1}} \otimes \hat{\sigma}_a \right) \\
	+ \sum_{a,b=x,y,z} \langle \hat{S}_{a} \otimes \hat{I}_{b} \rangle  \hat{\sigma}_a\otimes \hat{\sigma}_b .
\end{multline*}
Only the first line is fully expressed in the Bloch spheres, thus averaging out the entanglement terms with $\langle \hat{S}_{a} \otimes \hat{I}_{b} \rangle$.
It is not surprising then that by taking a partial trace of the whole system (Eq.~\ref{eq:rho_I}), the qubit subsystems are in mixed states.
More interestingly, this demonstrates that the initial population of the $\ket{0,0}$ state is distributed into the off-diagonal terms of the density matrix representing entanglement that are not explicitly pictured in the Bloch sphere.
And after a certain number of pulses, the off-diagonal terms evolve back into population terms, resulting in a population inversion to the $\ket{0,1}$ state.
This behavior is evidenced by the demonstrated full quantum state tomography with the 15 observables~\cite{QIP_NMR} for $N=0$ (initial state preparation) $N=6$ (intermediate `$\pi/2$' state) and $N=12$ (final `$\pi$' state) in Figs.~\ref{fig:DM_ibmq}~(b), (c) and (d) respectively.

At $N=0$, the IBMQ has a nearly perfect initialization of the qubits, with small deviations in some of the entanglement terms, likely due to readout errors related to the rotations necessary for the projection into the measurement axis and at $N=12$, we observe a complete population inversion of the target qubit.
By taking an intermediate number of pulses as $N=6$, we obtain a highly entangled state with concurrence 0.899~\cite{concurrence1, concurrence2}, close to a maximally entangled state with concurrence 1.
In this way, the DD-gates can potentially be employed as a tool for entanglement generation, apart from population changes to the target qubits, which motivates further research into this time-evolution dynamics.

\begin{figure*}[t!]
	\includegraphics[width=.2\textwidth]{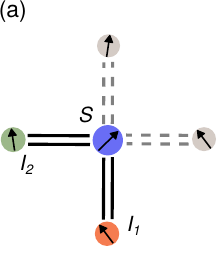}\hfil
	\includegraphics[width=.76\textwidth]{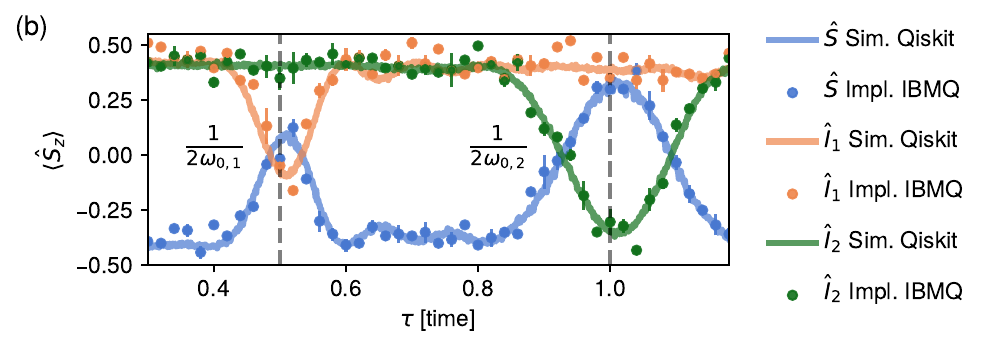}
	\includegraphics[width=.46\textwidth]{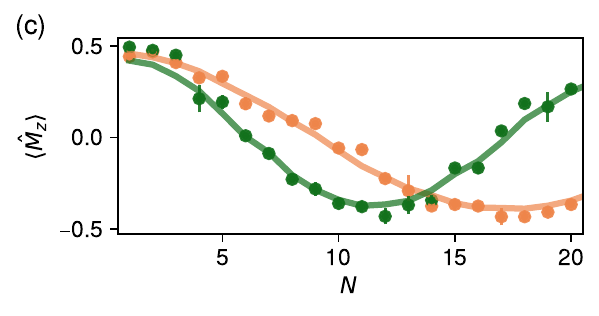}\hfil
	\includegraphics[width=.53\textwidth]{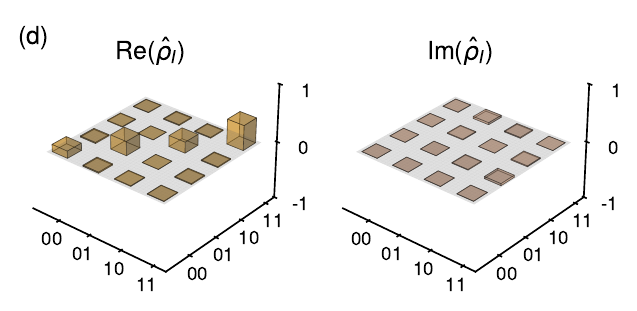}
\caption[DD-Gate Implementation with Two Target Qubits in IBMQ]{
		\textbf{(a)} Central qubit coupled to two target qubits represented with IBMQ.
		\textbf{(b)} Implemented and simulated CPMG-10 sequence applied to the central qubit coupled with two target qubits.
		The central qubit observable $\langle \hat{S}_z \rangle$ shows resonances at both of the target qubits resonant $\tau$ values, while each of the target qubits has its own resonance determined by $\tau=1/(2\omega_{0, j})$.
		\textbf{(c)} DD-mediated gates obtained with $\tau$ fixed to each qubit's resonance, showing a significant signal decay.
		\textbf{(d)} Implemented tomography of the reduced density matrix of the target qubits, with a partial trace over the control qubit.
		By applying a CPMG-16($\tau=0.5$) sequence followed by a CPMG-11($\tau=1.0$), both target qubits populations are inverted with a low fidelity of 0.659 with respect to the ideal state $\ket{11}$, while taking a partial trace of the control qubit.
		This poor performance of the DD-gate protocol with two  target qubits suggests a fundamental limitation for a digital gate-based quantum simulator to emulate a time-dependent Hamiltonian.
	}
	\label{fig:3qubit}
\end{figure*}

At the final state, small deviations from a perfect population inversion of the target qubit are also observed.
This can be attributed to Trotterization~\cite{trotterization1} and readout errors as seen at $N=0$.
But mainly, this discrepancy is related to the discrete nature of the $N$ variable, which does not allow us to take the exact moment during the time evolution where the state is completely inverted.
The demonstrated final state fidelity with the ideal $\ket{0,1}$ state is 0.945.
But if we disregard the control qubit (which serves more as an auxiliary qubit) and take the partial trace of the system (Eq.~\ref{eq:rho_I}), the fidelity for a population inversion of the target qubit is 0.995.
Another metric which is related to this fidelity is the amplitude of the Rabi-like oscillation, denoted as the pseudo-fidelity $\tilde{F}$.
For the implementation data in Fig.~\ref{fig:cpmg_ibmq}~(b), we have $\tilde{F}=0.963$.
This pseudo-fidelity has the advantage of not requiring a full quantum state tomography as performed here, thus being less computationally costly for IBMQ and which will be used in the remaining text.


\subsection{Two Target Qubits Gate}\label{sec:IBMQ_3}

In order to extend our model  to multiple target qubits, we perform DD-gate in a three-qubit system, composed of the central qubit coupled with two target qubits, as schematically represented in Fig.~\ref{fig:3qubit}~(a).
We consider a Hamiltonian as in Eq.~\ref{eq:IBMQ_H0_A}, with the same parameters for the control qubit as in Sec.~\ref{sec:NV_2qubit} and target qubits parameters of $A_{zx,1}=0.15$, $A_{zx,2}=0.10$, $\omega_{0, 1}=1.0$ and $\omega_{0, 2}=0.5$, all in units of [frequency].
In addition, we take a shorter time-step of $\Delta t = 0.0001$, in order to achieve better Trotterization approximation (Eq.~\ref{eq:Up}).

First, we experimentally measured and simulated the expectation values of the three qubits $\langle \hat{S}_z \rangle$, $\langle \hat{I}_{z,1} \rangle$ and $\langle \hat{I}_{z,1} \rangle$ under a CPMG-10 sequence, as shown in Fig.~\ref{fig:3qubit}~(b). 
The target qubits have resonance at their own Larmor frequencies $\tau=1/(2\omega_{0, j})$, without any response at a pulse separation resonant with the other qubit.
The central qubit has resonances at both frequencies, which leads to spectral ambiguities in quantum sensing applications with DD sequences~\cite{ambiguous_resonances, CROT_NV_DD}. 

To realize the DD-gate with the target qubits, the resonant pulse separation of each one is separately addressed with the CPMG sequence varying $N$, as previously done in Fig.~\ref{fig:cpmg_ibmq}~(b).
The experimental and simulation results are shown in Fig.~\ref{fig:3qubit}~(c), demonstrating full population inversions at $N=16$ in the target qubit 1 and $N=11$ in qubit 2.
The signal from both observables decays, evidencing a limitation of the IBMQ digital quantum simulator to reproduce the model for two target qubits.
These decays are also present in the Qiskit simulations, which validates the noise model employed in the simulations of the superconducting hardware.

The fidelities of each target-qubit DD-gate are obtained by an experimental quantum state tomography, see Fig.~\ref{fig:DM_ibmq}.
For the target qubit 1, the resulting fidelity with respect to the ideal state $\ket{10}$ is 0.702, where we take the partial trace of the control qubit, whereas in target qubit 2, we obtain 0.695 compared to the state $\ket{01}$.
Both fidelities are limited by the signal decay of the DD time-evolution.
Finally, the two DD-gates can be concatenated together, that is, being applied one after the other in the form of CPMG-16($\tau=0.5$)-CPMG-11($\tau=1.0$) in order to have a double population inversion of the two states yielding $\ket{11}$.
The obtained density matrix is shown in Fig.~\ref{fig:3qubit}~(d), with a fidelity of 0.659, again taking the partial trace of the control qubit.

These low fidelities are not necessarily limitations of the protocol, but they are are fundamentally limited by the intrinsic time-dependent nature of the Hamiltonian, which at the current technological stage cannot be properly executed using a discrete set of quantum gates within the IBMQ quantum simulator.
This does not imply, however, that the DD mediated gate is not suitable for multi-qubit systems.

\section{Experimentally Realistic Model with \textsuperscript{15}NV}\label{sec:NV}

Having introduced a general hardware-agnostic model for DD-mediated gates, tested on a digital quantum simulator, we now focus on an experimentally realistic case.
Among the various physical platforms utilized in quantum technologies, the NV center in diamond has been a focal point of research~\cite{NV_review1, NV_review2}, with applications ranging from quantum sensing~\cite{sensing_NV1, sensing_NV2, sensing_NV3}, to networks~\cite{communication_NV1} and memory-tokens~\cite{memory_NV, qtoken_1}.
The NV perfectly fits the description of a central qubit coupled to a target one, as developed in Sec.~\ref{sec:IBMQ}, and here we extend the model to it.
First, in Sec.~\ref{sec:H_NV}, we introduce the NV Hamiltonian and the simulation framework.
In Sec.~\ref{sec:NV_2qubit}, we present experimental measurements of the electron spin and simulations of the nuclear spin for the DD-mediated gate.
Finally, in Sec.~\ref{sec:pol_gen}, we discuss a possible application of the DD-gate for the polarization of the nuclear spins from an initial maximally mixed thermal state, supported by simulations.
Further details on the experimental realization with NVs can be found at Appendix~\ref{appendix:nv_exp}, while the simulations of the DD-gates with NVs are provided in the QuaCCAToo tutorials section~\cite{quaccatoo_git}.


\subsection{Theoretical Framework}\label{sec:H_NV}

In order to describe the time-evolution of the NV system under a DD sequence, we use a Hamiltonian model~\cite{NV_hamiltonian1, NV_hamiltonian2, NV_hamiltonian3} for the electron spin operator $\hat{\mathbf{S}}$ and the $^{15}$N nuclear spin $\hat{\mathbf{I}}$.
Taking the $z$-axis aligned with the NV crystal axis in the principal axis system (PAS), it can be written as
\begin{equation}\label{eq:H0_NV}
	\hat{H}_{0} = D \hat{S}_z^2
	- \gamma^e \mathbf{B}_0 \cdot \hat{\mathbf{S}}
	-  \gamma^n \mathbf{B}_0 \cdot \hat{\mathbf{I}} + \hat{\mathbf{S}} \cdot \mathbf{A} \cdot \hat{\mathbf{I}} .
\end{equation}
The first term represents a zero-field splitting due to the dipolar coupling between the two electrons forming the electron spin, with $D=2.87$~GHz.
The second and third terms are the Zeeman interactions with the external field $\mathbf{B}_0 $, where the gyromagnetic ratios of the two spins are $\gamma^e=-28.025$~GHz/T and $\gamma^n = -4.316$~MHz/T.
At magnetic fields below $|\mathbf{B}_0|<102.4$~mT, the $D$ term is dominant over the electron Zeeman one and the lowest energy state corresponds to the $m_S=0$ level.
The fourth term corresponds to the hyperfine interaction between the two spins, which in the PAS frame is represented by a diagonal coupling tensor $\mathbf{A}$ with $A_{xx}=A_{yy}=3.65$~MHz and $A_{zz}=3.03$~MHz~\cite{Newton09}.

In most applications it is assumed that $\mathbf{B}_0 \parallel z $, but as recently shown~\cite{ambiguous_resonances}, a small misalignment induces $\hat{S}_x$ and $\hat{I}_x$ terms in the Zeeman interactions, which lead to a detection of the $^{15}$N nuclear spin precession by the electron spin in DD sequences.
In the context of quantum sensing, this effect results in ambiguous resonances in the DD spectrum, which can be confused with the signal intended to be measured.
Here, these resonances are the means for the DD-mediated nuclear gates with the $^{15}$NV system.
When considering the $^{14}$N isotope on the other hand, the quadrupole interaction fixes the nuclear spin, suppressing its precession and making the application of such DD-gates less practical.
The NV Hamiltonian under these considerations does not exactly correspond to the minimal two-qubit Hamiltonian developed in Sec.~\ref{sec:IBMQ_H0} (Eq.~\ref{eq:ibmq_H0_2q}), since the Zeeman interactions have perpendicular terms with $\hat{S}_x$ and $\hat{I}_x$ instead of the hyperfine coupling.
However, with a change of the reference frame as discussed in Appendix~\ref{appendix:NV_PAS}, we see that both cases are the physical manifestation of the same effect.

Prior to the application of the DD sequences, the electron spin of the NV is initialized into the state $\ket{0} \equiv \ket{m_S=0}$ by an optical pumping mechanism~\cite{optical_pumping} using a green laser.
In general, the $^{15}$N nuclear state is in an initial mixed thermal state, with a density matrix practically equal to identity $\hat{\mathds{1}}/2$~\cite{quaccatoo_paper}, unlike in the IBMQ.
The nuclear spin can be polarized through projective measurements via the electron spin~\cite{single_shot_readout}, yet this requires precise alignment of the $\mathbf{B}_0$ field along the NV axis, which would be incompatible with the experimental design of the DD-gate, since the Zeeman terms along $x$ are required.
Overall then, the state of the NV system is described by mixed density matrices, rather than pure ket states.

The control field interaction with the NV system, given by $\hat{H}_1(t)$, is analogous to Eq.~\ref{eq:IBMQ_H1}.
With the only difference being that due to the spin multiplicity $S=1$, we select the $m_S=0\rightarrow-1$ transition frequency for the control pulse, with $\ket{1} \equiv \ket{m_S=-1}$ forming then the computational basis.
The excitation bandwidth has to be smaller than the energy splitting between $m_S=-1$ and $m_S=+1$, in order to avoid double quantum excitations of the two electron levels~\cite{double_quantum}.
If this condition is met, the three-level system can be effectively treated as a qubit.

When considering the interaction of the NV's electron spin and surrounding $^{13}$C nuclei, the hyperfine term is typically smaller or in the same order of the nuclear Zeeman interaction.
This permits to calculate the DD resonance features in an analytical framework, based on the adoption of rotating frames and assumption of RWA~\cite{CROT_NV_DD, CROT_13C_3, CROT_13C_4, CROT_13C_5, CROT_13C_6}.
For this case however, with the $^{15}$N nuclear spin, the hyperfine term is dominant over the nuclear Zeeman interaction for moderate magnetic fields  $|\gamma^n \mathbf{B}_0| < |A_{xx}|, |A_{zz}|$ and the break of symmetry of the misaligned $\mathbf{B}_0$ introduces non-commuting terms with the rotation operators of the rotating frame~\cite{ramsey_NV}, making an analytical solution unfeasible.
The dynamics of the system under the $\hat{H}_0 + \hat{H}_1(t)$ then needs to be solved in the static laboratory frame with the Liouville-von Neumann equation~\cite{QME}, motivating the choice of this frame in Sec.~\ref{sec:IBMQ_H0}.
To numerically simulate this dynamics, we use the Quantum Color Centers Analysis Toolbox (QuaCCAToo)~\cite{quaccatoo_git, quaccatoo_paper} software, based on QuTip~\cite{qutip}, which is a more complete tool for simulating color-center systems than Qiskit.

At the end of the pulse sequence, we measure the fluorescence observable given by $\langle \hat{F} \rangle \equiv \text{Tr} (2 \hat{\rho}_0 \hat{\rho}_f)$~\cite{quaccatoo_paper}, where $\hat{\rho}_f$ is the final state of the system.
With this definition, $\langle \hat{F} \rangle $ is related but not strictly the same as the expectation value of the $\hat{S}_z$ operator, as the $m_S=\pm1$ states occupation cannot be distinguished by the emitted fluorescence.
But since we are exciting just one of the $m_S$ levels, both coincide and we refer to $\langle \hat{F} \rangle $ simply as the expectation value of the spin operator $\langle \hat{S}_z \rangle$.
As for the $^{15}$N nuclear spin observable, we use the $\langle \hat{I}_z \rangle$ observable as in Sec.~\ref{sec:IBMQ_H0}.
Noting, however, that $\hat{S}_z$ has expectation values from 0 to 1 in this definition, while $\hat{I}_z$ has from -0.5 to 0.5.


\subsection{Realization of \textsuperscript{15}NV DD-Gate}\label{sec:NV_2qubit}

In order to characterize the DD resonance spectrum of a $^{15}$NV in a misaligned magnetic field, a single NV was measured under a magnetic field of $|\mathbf{B}_0|=32$~mT at misalignment angle of $\theta_0=2.9^\circ$.
Here we use the XYN sequence instead of the CPMG, due to its better pulse errors resilience, as demonstrated in Appendix~\ref{appendix:pulse_errors}.
The XYN is similar to the CPMG, being composed by $N$ $\pi$-pulses plus the initial and final projective $\pi/2$-pulses, but with the $\pi$-pulses intercalated between the $x$ and $y$ axes~\cite{XY8, DD_NV}.
Further experimental details are given in Appendix~\ref{appendix:nv_exp}.

\begin{figure}[t!]
	\includegraphics[width=\columnwidth]{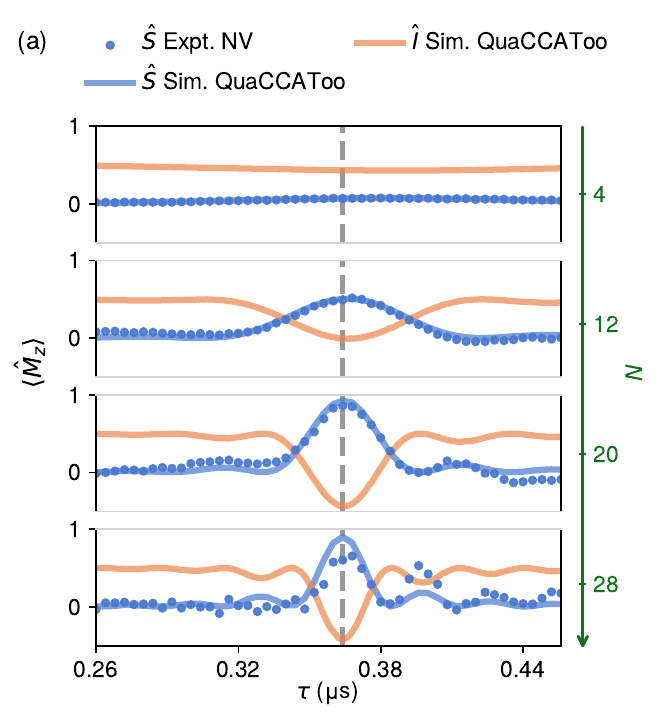}\vspace{-.3cm}
	\begin{flushleft}
		\includegraphics[width=.85\columnwidth]{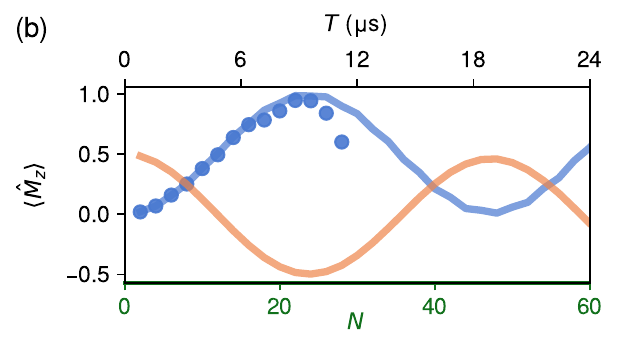}
	\end{flushleft}
\caption[Experimental Realization of DD-Gate with \textsuperscript{15}NV]{
		\textbf{(a)} XYN spectra of $^{15}$NV for different number of pulses $N$.
		A small misalignment of $\mathbf{B}_0$ causes the precession of the $^{15}$N nuclear spin to be sensed by the electron spin observable $\langle \hat{S}_z \rangle $, experimentally measured and simulated.
		Although the experimental measurement of the $\langle \hat{I}_z \rangle$ observable was not possible in this work, its simulations indicate resonances opposite to $\langle \hat{S}_z \rangle$.
		As $N$ increases, the experimental data starts to deviate from the simulations, due to decoherence and pulse errors~\cite{ambiguous_resonances}.
		\textbf{(b)} DD-gate with $^{15}$NV.
By taking $\tau=0.364$~\textmu s fixed and varying $N$, we observe Rabi-like oscillations of the two observables.
With 24 pulses, a complete inversion of the population is observed, totaling $T_\pi=9.04$~\textmu s, which can represent a speed-up of nearly 10 times compared to direct excitation of the $^{15}$N nuclear spin.}
	\label{fig:XYN_NV}
\end{figure}

The experimentally measured and simulated expectation values of $\langle \hat{S}_z \rangle$ for different values of pulses $N$ are shown in Fig.~\ref{fig:XYN_NV}~(a).
The remaining values of $N$ can be found at~\cite{quaccatoo_git}.
For comparison, we also show the simulated data for $\langle \hat{I}_z \rangle$.
While the direct experimental measurement of the nuclear spin polarization was not feasible in this work, the simulations provide valuable insights into the underlying dynamics of the system.
In the simulations, we assume an initially polarized state $\ket{00}=\ket{m_S=0}\otimes\ket{m_I=+1/2}$, while the initial state in the experiments is a mixed state corresponding to $\ket{m_S=0}\bra{m_S=0}\otimes\hat{\mathds{1}}/2$~\cite{quaccatoo_paper}.
This approximation is justified, because the control spin observable is not affected by the initial state of the target spin under the XYN sequence, as observed in Sec.~\ref{sec:pol_gen}.

The simulation and experimental values of $\langle \hat{S}_z \rangle$ have a prominent resonance from the $^{15}$N nuclei at around $\tau=0.364$~\textmu s, in accordance with Ref.~\cite{ambiguous_resonances}.
In contrast to IBMQ, here the resonant pulse separation does not correspond to $\tau=1/(2|\gamma^n \mathbf{B}_0|)$, given that the hyperfine coupling is dominant over the Zeeman term $|A_{zz}|,|A_{xx}|>|\gamma^n \mathbf{B}_0|$ and the nuclear spin has a different Larmor frequency for each electron level.
Moreover, the simulations of the nuclear spin observable $\langle \hat{I}_z \rangle$ indicate that its time-evolution is opposite to $\langle \hat{S}_z \rangle$, as seen in Sec.~\ref{sec:IBMQ} with the IBMQ model.

The intensity of the resonances again changes with number of pulses, where the DD mediated gate can be obtained by taking $\tau=0.364$~\textmu s fixed and varying $N$ [Fig.~\ref{fig:XYN_NV}~(b)].
Thus, the expectation values of $\langle \hat{S}_z \rangle$ and $\langle \hat{I}_z \rangle$ perform the Rabi-like oscillations analogous to Fig.~\ref{fig:cpmg_ibmq}~(b), with a complete inversion of their populations at 24 pulses, totaling $T_\pi=9.04$~\textmu s.
This value is close to 100 times faster than the value reported in Ref.~\cite{CROT_NV_DD} for a $^{13}$C nuclear spin, given the much stronger hyperfine interaction with $^{15}$N, and almost 10 times faster than direct RF pulse excitation, assuming the same driving field intensity for the MW and RF fields (Appendix~\ref{appendix:nv_exp}).
This way, the DD-mediated gate does not only show potential for substantial short gate time, but it also reaches near unity pseudo-fidelity of $\tilde{F}=0.980$ for a complete population inversion of the $^{15}$N nuclear spin, as demonstrated by the simulations of $\langle \hat{I}_z \rangle$.
Still, the simulations of the nuclear spin observable are open to experimental verification.

As the number of pulses $N$ increases, decoherence and pulse errors affect the experimental values of $\langle \hat{S}_z \rangle$, leading to a deviation from the simulation, as similarly observed in other DD sequences under pulse errors~\cite{ambiguous_resonances}.
This confirms that the NV system is more susceptible to experimental errors than the IBMQ quantum simulator.
Nonetheless, the effect from pulse errors can be mitigated by pulse engineering techniques (Appendix~\ref{appendix:pulse_errors}) and the coherence of NV centers can also be extended by optimization of its fabrication methods~\cite{token_fabrication1, token_fabrication2}.
Independent of the sequence or system, QuaCCAToo~\cite{quaccatoo_git} provides the means for users to simulate these DD-gates to arbitrary color centers. 


\subsection{Polarization of the Target Nuclear Spin}\label{sec:pol_gen}

\begin{figure}[b!]
	\includegraphics[width=\columnwidth]{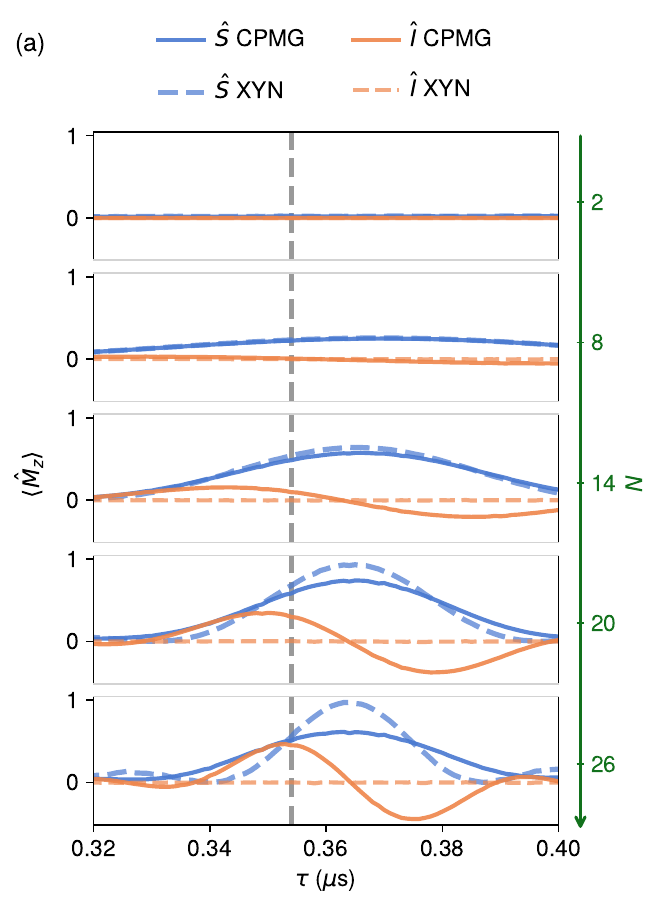}
	\includegraphics[width=\columnwidth]{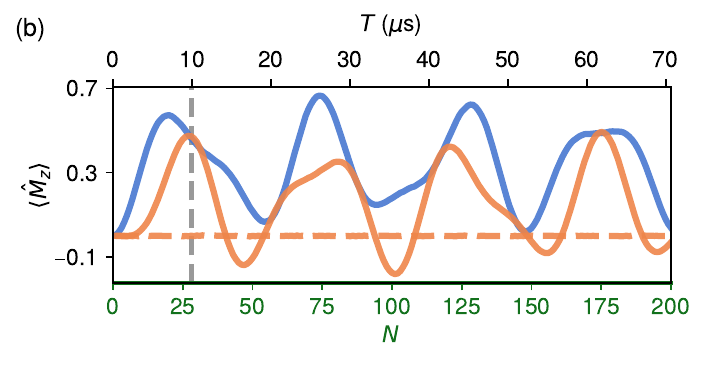}
	\caption[Simulation of Polarization Generation in Target \textsuperscript{15}N Nuclear Spin]{
	\textbf{(a)} Simulated $^{15}$NV observables under CPMG and XYN sequences for different number of pulses $N$, with an initial state $\hat{\rho}_0 = \ket{0}\bra{0}\otimes\hat{\mathds{1}}/2$.
	In the XYN sequence, $\langle \hat{S}_z \rangle$ is invariant with the initial state of the target qubit, while $\langle \hat{I}_z \rangle$ remains constant.
	Contrastingly, the CPMG sequence induces a completely different dynamics to the initially mixed nuclear state, showing an antisymmetrical resonance around $\tau=0.364$~\textmu s.
	\textbf{(b)} Simulated polarization of the target qubit by DD-gate.
	By keeping $\tau=0.3534$~\textmu s fixed and varying $N$, the  expectation values of the two observables show anharmonic oscillations, where with $N=27$ pulses totaling $T_{pol}=9.54$~\textmu s, $\langle \hat{I}_z \rangle$ reaches its maximum value corresponding to a polarization efficiency of 0.950, while the electron $\langle \hat{S}_z \rangle$ is perfectly mixed.
	This way, the initial polarization of the electron spin is transferred to the nuclear one, mediated through entanglement terms.
	}
	\label{fig:NV_polarization}
\end{figure}

A valuable application of this DD-gate is for the polarization of the target qubits.
At room temperature and typically employed magnetic fields with intensities in the orders of a few mT up to hundreds of mT, the nuclear spins coupled to the NV are nearly in a maximally mixed state.
This thermal polarization given by the Boltzmann distribution is hindering many quantum technology protocols and solutions for polarizing the spins coupled to the NV are highly valued.
Different methods have been developed for polarization of
the $^{15}$N nucleus~\cite{single_shot_readout} coupled to the NV and even spin ensembles~\cite{DNP1}, based on the Hartman-Hahn condition~\cite{HH1} or with complex Hamiltonian engineering sequences~\cite{DNP2}.
Here, we propose a new method with less restrictive experimental constraints, based on simple dynamical decoupling sequences, without requiring RF pulses to the nuclear spin, precise magnetic-field alignment, intricate pulse generation, or long-coherence NV diamond samples.

The pulse error resilience analysis performed in Appendix~\ref{appendix:pulse_errors} indicates that the CPMG and XYN result in different resonant responses from the target qubit.
Yet, they also behave differently depending on the initial state of the target qubit.
To show that, Fig.~\ref{fig:NV_polarization}~(a) presents simulations of the NV spectrum under CPMG and XYN sequences considering an initial state $\ket{0}\bra{0}\otimes\hat{\mathds{1}}/2$.
The number of pulses $N$ is varied, while considering the same experimental parameters as in Sec.~\ref{sec:NV_2qubit}, apart from this initial state.
Firstly, the electron spin under the XYN sequence gives the same resonances as observed in Fig.~\ref{fig:XYN_NV}~(a) when considering the initial state $\ket{m_S=0}\otimes\ket{m_I=+1/2}$.
This shows that $\langle \hat{S}_z \rangle$ does not depend on the initial state of the nuclear spin in the XYN sequence.
By comparing the central spin resonance between the two sequences, we also observe that XYN generates narrower linewidths and more intense resonance peaks than CPMG.

The most prominent difference between the two sequences, nonetheless, arises from the comparison of the coupled nuclear spin response in each sequence.
In the XYN sequence, the nuclear spin observable shows no variation, maintaining a constant value of $\langle \hat{I}_z \rangle=0$, a behavior that is somewhat intuitive for a qubit that undergoes a unitary evolution from an initial state $\hat{\mathds{1}}/2$.
On the other hand, the CPMG induces variations in the $\langle \hat{I}_z \rangle$ observable even for an identity initial state, being 0 at the resonant $\tau$ and anti-symmetrical around this point.
This is not violating the unitarity of the time-evolution, because as shown in Fig.~\ref{fig:DM_ibmq}~(a) and discussed in detail in Sec.~\ref{sec:IBMQ_2}, the DD sequence is able to alter the Bloch sphere radius (the mixing) of the coupled qubit subsystem, obtained from a partial trace of the whole system (Eq.~\ref{eq:rho_I}).
In contrast to what is performed with IBMQ, it is then possible to obtain a polarized state from an initially mixed one. 
Altogether, this indicates that by simply intercalating the axis of the periodical reversal in the DD sequences, we obtain completely different dynamics from the CPMG and XYN sequences, resulting from variations in their filter function.
This difference between both sequences goes beyond mere pulse errors resilience.

To further characterize this effect of polarization, we simulate the time evolutions of both observables in the CPMG sequence for fixed pulse separation $\tau=0.3534$~\textmu s and varying $N$.
For comparison, we also show the nuclear spin observable in the XYN, which remains nearly constant even for a large number of pulses.
Both observables in the CPMG undergo anharmonic oscillations, that continue indefinitely without pulse errors, decoherence or relaxation.
By taking $N=27$ with $T_{pol}=9.54$~\textmu s in the CPMG sequence, we have a maximum of $\langle \hat{I}_z \rangle = 0.475$, representing an efficiency of 0.950 compared to the maximum possible value of $\langle \hat{I}_z \rangle = 0.5$.
At this value of $\tau$, the electron spin observable is also close to 0.5, half of its maximum value of 1.
This duration of the polarization sequence can also represent a considerable speed-up as compared to more elaborate Hamiltonian-engineering sequences~\cite{DNP2}, being over 3 times faster than the value reported for a NOVEL sequence and nearly 5 times faster than PulsedPol~\cite{DNP2}.

In order to demonstrate that the final state of the nuclear spin is actually polarized, we simulate a quantum state tomography, as presented in the Appendix (Fig.~\ref{fig:NV_DM}).
This demonstrates that the final state is $\hat{\mathds{1}}/2 \otimes \ket{m_I=+1/2}\bra{m_I=+1/2}$, with a fidelity of 0.979.
The fact that the electron spin is in a mixed state at the end of the DD-gate confirms that its initial polarization is transferred to the coupled qubit.
If we instead take the antisymmetrical $\tau$ point above the resonance where $\langle \hat{I}_z \rangle$ is negative, we are able to polarize the nuclear spin into the other state.
Altogether, this suggests the possibility for a high-fidelity polarization generation through a simple CPMG sequence on the NV central spin.
Clearly, this prediction also requires future experimental validation.

\section{Conclusion and Outlook}\label{sec:conclusion}

 High-fidelity gates are crucial for quantum information processing.
In this work, we provide a general framework for the implementation of such gates through DD of a central qubit coupled to target qubits.
The critical role of DD is emphasized, enabling precise, fast, and high-fidelity control of the target qubits by mitigating environmental noise and eliminating the need for long, error-prone resonant control of the target qubits.
The DD-mediated gates are thoroughly analyzed and implemented with IBMQ and NVs, supported by numerical simulations within the two different frameworks.

Initially, we develop a general model with minimal hardware assumptions and benchmark it on IBMQ (Sec.~\ref{sec:IBMQ}), where rotating frames and other perturbative approximations~\cite{ramsey_NV, beyond_RWA} were not used. This allows our model to operate with arbitrary interactions (Appendix~\ref{appendix:Azx}) that do not necessarily commute with the rotation operators of the rotating frame.
The presence of a coupling between $\hat{S}_z$ (control qubit) and $\hat{I}_x$ of the target qubit results in resonances in the DD spectrum (Sec.~\ref{sec:IBMQ_2}), which can be leveraged to obtain state transitions of the target qubit without its direct control by resonant excitation.
Upon further analysis of the time-evolution, we observe that the DD-gate does not correspond to proper qubit rotations, but instead, the qubits evolve from one state to another through points inside the Bloch sphere, mediated via entanglement terms of the full system density matrix.
In addition, the solution of the qubit dynamics in the static laboratory frame straightforwardly incorporates pulse errors within the control field (Appendix~\ref{appendix:pulse_errors}).

While we demonstrate near-unity fidelities for the DD-gate with one target qubit in IBMQ, the implementation with two target qubits results in lower fidelities (Sec.~\ref{sec:IBMQ_3}).
This significant reduction in performance does not indicate a limitation of the DD-gate method itself but rather a more fundamental challenge of the current NISQ era~\cite{nisq, nisq2fasq} hardware, namely the emulation of a time-dependent Hamiltonian in a digital gate-based architecture, as provided by IBMQ.
This emphasizes the necessity for specialized analog quantum simulators~\cite{qsimulator}, in which time-dependent Hamiltonians can be physically implemented in a direct way, where errors introduced by Trotterization~\cite{trotterization1, lie_trotter} and compilations are avoided.

Overall, we observe excellent agreement between the theoretical model and the IBMQ implementation, which demonstrates the hardware agnosticism of the DD-gate method.
The latter can be implemented in other physical systems, as long as they are consistent with the central-target qubit description.
Notably, our gate method is shown to be particularly well-suited for color centers.
By extending the minimal model to $^{15}$NVs (Sec.~\ref{sec:H_NV}), we show that the electron spin can be used as the control qubit to induce a DD-gate on the $^{15}$N nuclear spin.
Experimental measurements of the electron spin and QuaCCAToo simulations~\cite{quaccatoo_git, quaccatoo_paper} of the nuclear spin indicate the possibility of high-fidelity control of the target qubit, with substantial speed-ups in gate duration as compared to direct excitation or previous DD-mediated methods~\cite{CROT_NV_DD}.
The experimental implementation with NVs also highlights the sensitivity of the protocol to pulse imperfections, where for longer pulse sequences we observe discrepancies with the simulation model.
These errors could be mitigated by employing advanced pulse sequences~\cite{IBMQ_pulse_errors, UDD, RXY8_1, RXY8_2, RXY8_g}.

Another possible application of the DD-gates is the polarization of target qubits, which are commonly in mixed thermal states.
Here, we propose and simulate a high-efficiency method with NVs using a simple CPMG sequence.
This could represent a significant simplification of the experimental conditions of current methods for polarization, where precise magnetic field alignment~\cite{single_shot_readout} or intricate pulse sequences~\cite{DNP1} based on Hamiltonian engineering~\cite{DNP2} are not required.
The natural progression of this work is then the experimental verification of the proposed polarization protocol and the measurement of the $^{15}$NV nuclear spin observable (Sec.~\ref{sec:NV_2qubit}), in order to fully validate the DD-gate application with NVs.

Further generalizations of the model can be obtained by introducing interactions between the target qubits, which are neglected in this work.
However, these could potentially be leveraged by the DD technique as well, in order to obtain even more robust multi-qubit DD-gates.
This study also focuses on only two possible initial states for the target qubit: $\ket{0}$ and $\hat{\mathds{1}}/2$, representing a part of a broader operational landscape.
Moreover, the method shows potential for the generation of deterministic entangled states with high concurrence (Sec.~\ref{sec:IBMQ_2}), which is not analyzed in depth here.
Future studies can then characterize the complete quantum channel induced by the DD-gates, enabling its utilization in general quantum algorithm applications.

This work also provides a critical step towards practical quantum technology implementations with NVs.
Scalability is paramount for applications such as the diamond-based quantum token~\cite{qtoken_1, qtoken_2}, which requires identical operation across different NVs.
By using the intrinsic $^{15}$N spin instead of stochastic $^{13}$C nuclei as in previous studies~\cite{CROT_13C_1, CROT_13C_2, CROT_13C_3, CROT_13C_4, CROT_13C_5, CROT_13C_6}, we transition from a system limited by random lattice positions to one governed by deterministic interactions.
Last but not least, the open-source simulations and implementations with IBMQ and NVs are themselves primary contributions of this work, providing a viable framework for quantum simulation of time-dependent qubit dynamics in the laboratory frame within NISQ-era digital gate-based quantum simulators.

\section*{Data and Code Availability}

All implementation and simulation code used in this work is open source.
Qiskit implementations and simulations with IBMQ are provided in the author's Github repository~\cite{dd_gates_git}.
Simulations regarding NVs can be found at the QuaCCAToo documentation~\cite{quaccatoo_git}.
The data used in this work is available for scientific use upon reasonable request.

\section*{Author Contributions}

L.T. conceptualized this work, developed the theory, performed experiments and simulations with IBMQ and NV center.
M.D. assisted in the experiments with NV center.
K.V. developed and maintained the experimental setup for the experiments with NV center, and participated in the development of QuaCCAToo.
B.N. acquired funding and supervised the work.
All authors contributed to the writing of the text.

\section*{Acknowledgments}

We are grateful to Dr. Tommaso Pregnolato, Prof. Dr. Tim Schröder, Alexander Külberg and Dr. Andreas Thies from Ferdinand-Braun-Institute (FBH) for performing the ion implantation in the diamond sample.
We acknowledge Miriam Mendoza Delgado and Prof. Dr. Cyril Popov from the Institute of Nanostructure Technologies and Analytics of the University of Kassel for the annealing of the diamond sample.
We also thank Sergei Trofimov and Anmol Singh for their contribution to the QuaCCAToo software.
This work was supported by the \textit{Bundesministerium für Bildung und Forschung} (BMBF) under the project  \textit{DIamant-basiert QuantenTOKen} (DIQTOK - n\textsuperscript{o} 16KISQ035) and \textit{Diamant-basierte Quantenmaterialien} (DIAQUAM - n\textsuperscript{o} 13N16956).
In addition, this work received funding from the \textit{Deutsche Forschungsgemeinschaft} (DFG) grants 410866378 and 41086656.

\appendix

\section{Implementation and Simulation Methods}

\subsection{IBMQ and Qiskit}\label{appendix:qiskit}

All implementations on IBMQ were performed using the Qiskit software development kit (SDK)~\cite{qiskit}, version 2.1.0, and executed on the Torino quantum processing unit via the Qiskit Runtime architecture, version 0.40.1.
The simulations of IBMQ with realistic noise models were executed locally using the Qiskit Aer simulator, version 0.17.1.
Given the fast development of the Qiskit software ecosystem, the codes provided at~\cite{dd_gates_git} and used in this work may be outdated.
The Hamiltonians (Eqs.~\ref{eq:IBMQ_H0_A} and~\ref{eq:IBMQ_H1}) are constructed based on the sparse Pauli operators representation from Qiskit.
These are then used to create the time-evolution operators (Eqs.~\ref{eq:Utau} and~\ref{eq:Up}) with Pauli evolution gates, synthesized via Lie Trotterization with one repetition~\cite{lie_trotter}.
Each Pauli evolution gate is appended to the quantum circuit respecting time-ordering, with discretized times during pulses (Sec.~\ref{sec:IBMQ_H0}).
Measurements were performed with 100 shots, except for the Bloch sphere tomography in Fig~\ref{fig:DM_ibmq}~(a), which was run with 500 shots for a better state reconstruction.

For transpiling the high-level circuit representations to native hardware gates, we used optimization level 2, with inverse cancellation of gates but no commutative cancellations.
The dense layout method is used, taking the qubits with longest coherence and smallest gate times.
Lastly, for translation, the gates are synthesized without any approximation.
The transpiled circuits are then added with the mapped observables to the primitive unified blocs list and executed in the IBMQ or simulated with Aer.
All demonstrations with IBMQ Torino were executed between July and August 2025, with minor fluctuations in some of the hardware parameters, as evidenced by the calibration reports.
In any case, this does not affect the results presented here.


\subsection{NV and QuaCCAToo}\label{appendix:nv_exp}

The NV center used in this work was fabricated in an electronic grade diamond plate (Element Six UK Ltd.) via  $^{15}$N ion implantation with energies between 20 and 30~keV and a dose of $10^{9}$~$^{15}$N/cm$^2$.
After implantation, the sample was annealed in ultra-high vacuum at 1000$^\circ$C.
The experimental measurements with the NV center were performed on a custom-built confocal microscope, as described in detail in Refs.~\cite{confocal_setup, ambiguous_resonances}.
In short, the external magnetic field $|\mathbf{B}_0|$ is applied using a permanent magnet, with four degrees of freedom.
The misalignment angle and field intensity is obtained from fits of the optically detected magnetic resonance spectrum~\cite{misalignement}.
The NVs are initialized and measured with a diode laser of wavelength 518~nm, focused on the NV with an air objective.
The photons are then detected and counted, giving the fluorescence observable.
MW pulses for central-spin control are generated by an arbitrary waveform generator, then amplified and transmitted to the diamond through a thin copper wire ($d\approx20$~\textmu m), resulting in a $\pi$-pulse duration of $t_\pi^{MW}=12.85$~ns, determined from Rabi oscillations~\cite{rabi}.
For this value of pulse-duration, the energy difference between the $m_S=\pm1$ levels is much larger than the bandwidth of the pulse and the system can be well approximated as a two-level system, even though QuaCCAToo simulations do not rely on this approximation.
If we assume the same control field intensity $B_1$ for a hypothetical RF pulse as for the MW pulse, this would lead to a direct $\pi$-pulse to the nuclear spin of duration $t_\pi^{RF}= | \gamma^e / \gamma_n | t_\pi^{MW} = 83.4$~\textmu s, as adopted in Sec.~\ref{sec:NV_2qubit}.  
MW $\pi_y$/2-pulses are included before and after the sequence to rotate the electron spin from and to the quantization axis, or in the case that the number of pulse blocks $N/2$ is odd, the final pulse is 3$\pi$/2 to preserve the sign of $\langle \hat{S}_z \rangle$.
All experimental control and data acquisition are done through the Qudi software~\cite{qudi}.

QuaCCAToo version 1.1.0 was used to simulate the NV dynamics in the DD-gates.
The \texttt{NV} and \texttt{XY} pre-defined classes are used to model the NV systems and the XYN sequence respectively, with the same parameters as experimentally measured.
The experimental data processing is done through the \texttt{ExpData} class, where a baseline correction of order two is applied to all XYN spectra, as shown in the repository~\cite{quaccatoo_git}.


\section{Coupling Factor}\label{appendix:Azx}

\begin{figure}[b!]
	\includegraphics[width=\columnwidth]{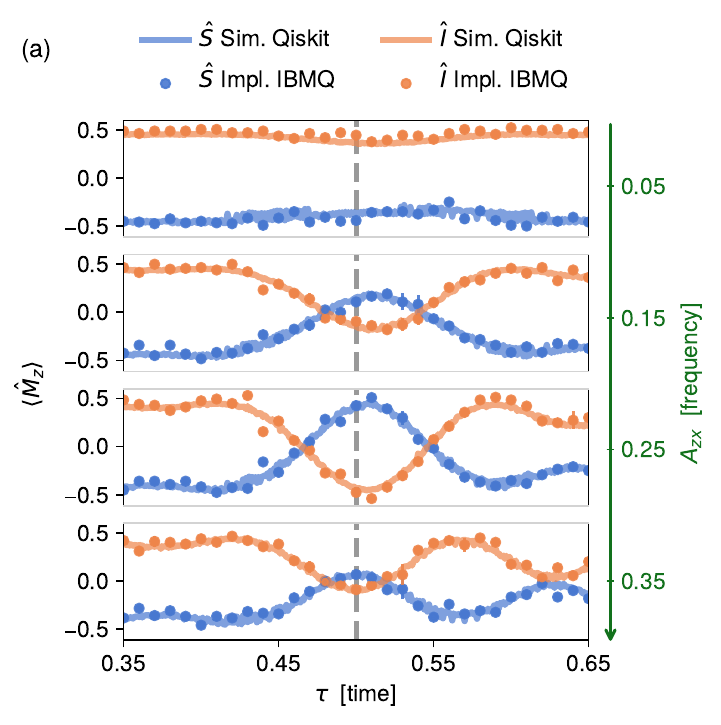}
	\vspace{.5cm}
	\includegraphics[width=\columnwidth]{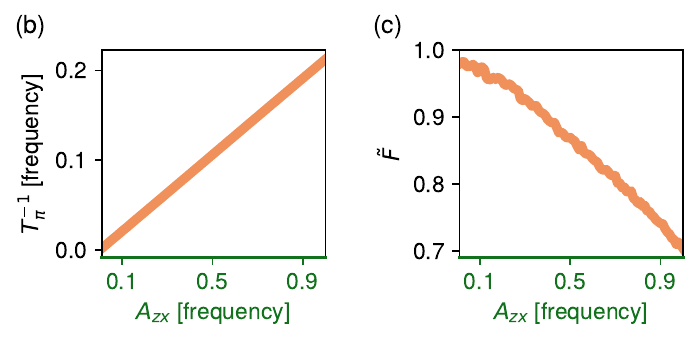}
	\caption[Influence from Coupling Factor $A_{zx}$ on DD-Gate with IBMQ]{
		\textbf{(a)} Implemented and simulated CPMG-10 sequences in IBMQ for different values of coupling factor $A_{zx}$.
		As $A_{zx}$ increases, the resonance amplitude changes and the sidebands become more pronounced.
		\textbf{(b)} Simulated frequency of the DD-gate $T_\pi^{-1}$ as a function of $A_{zx}$.
		$T_\pi^{-1}$ is linearly proportional to $A_{zx}$, showing that the DD-gate becomes faster with the coupling factor.
		\textbf{(c)} Simulated pseudo-fidelity of the DD-gate $\tilde{F}$ as a function of $A_{zx}$.
		$\tilde{F}$ experiences a non-linear decrease with $A_{zx}$, as with a faster DD-gate, the discrete pulse number parameter $N$ is less likely to determine the exact time for the population inversion of the target qubit.
	}
	\label{fig:cpmg_ibmq_Azx}
\end{figure}

The coupling factor $A_{zx}$ determines the resonant response of the central and target qubits.
To characterize its effect on the DD-gates, we consider the same conditions for the IBMQ as in Sec.~\ref{sec:IBMQ_2}, with a Hamiltonian equal to Eq.~\ref{eq:ibmq_H0_2q}, but with varying $A_{zx}$ between 0.05 and 0.35 [frequency].
In this range of coupling factors, the relation $\omega_{0, 0} \gg A_{zx}$ is still valid, meaning that the driving of the qubit is well-defined and harmonic.
With that, the CPMG-10 sequence was measured and simulated, as presented in Fig.~\ref{fig:cpmg_ibmq_Azx}~(a).
As the coupling factor increases, the resonance amplitude varies and the sidebands also get more pronounced, as similarly observed when $N$ is changed [Fig.~\ref{fig:cpmg_ibmq}~(a)].

The relation between $A_{zx}$ and the time required for a full population inversion of the target qubit $T_\pi$ is particularly interesting for the design of the DD-gates, as well as the resulting pseudo-fidelity of the gate $\tilde{F}$.
To obtain these values, we simulate the CPMG sequences with fixed $\tau=0.5$~[time] and different number of pulses $N$.
This leads to a series of Rabi-like oscillations as in Fig.~\ref{fig:cpmg_ibmq}~(b), which we fit with trigonometric functions, thus obtaining the $T_\pi$ and $\tilde{F}$ values as a function of $A_{zx}$.
Fig.~\ref{fig:cpmg_ibmq_Azx}~(b) shows that the frequency of the DD-gates is linearly proportional to the coupling factor $T_\pi^{-1} \propto A_{zx}$, following the relation $T_\pi \cong 0.212/A_{zx}$ for these Hamiltonian parameters.
However, it is important to note that $T_\pi$ is also influenced by other pertinent experimental parameters, such as the Larmor frequency of the target qubit.
Previous models based on perturbative approximations that do not consider these experimental parameters~\cite{spurious, multipulse1} would lead to a relation of $T_\pi = 1/(2 A_{zx})$, which underestimates the actual duration for a $\pi$ pulse by over a factor of 2.

It is clearly advantageous to have a strong coupling term for having short gates, but on the other hand, the fast driving of the target qubit results in a lower pseudo-fidelity $\tilde{F}$ for a full population inversion, as indicated in Fig.~\ref{fig:cpmg_ibmq_Azx}~(c).
At very weak couplings around $A_{zx}=0.1$ [frequency], the pseudo-fidelity reaches values close to 1, which then decays non-linearly to below 0.7 at $A_{zx}=1$.
This can be attributed to the discretized nature of the pulse number variable $N$, that with a faster oscillation is less likely to correctly determine the exact point in time of the full population inversion of the target qubit.


\section{Pulse Errors Resilience}\label{appendix:pulse_errors}

\begin{figure*}[t!]
	\includegraphics[width=.49\textwidth]{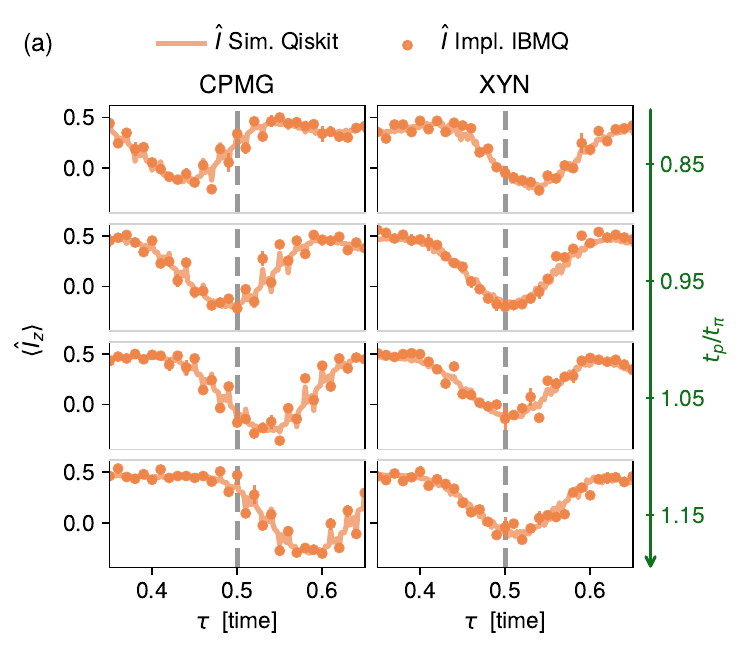}\hfil
	\includegraphics[width=.49\textwidth]{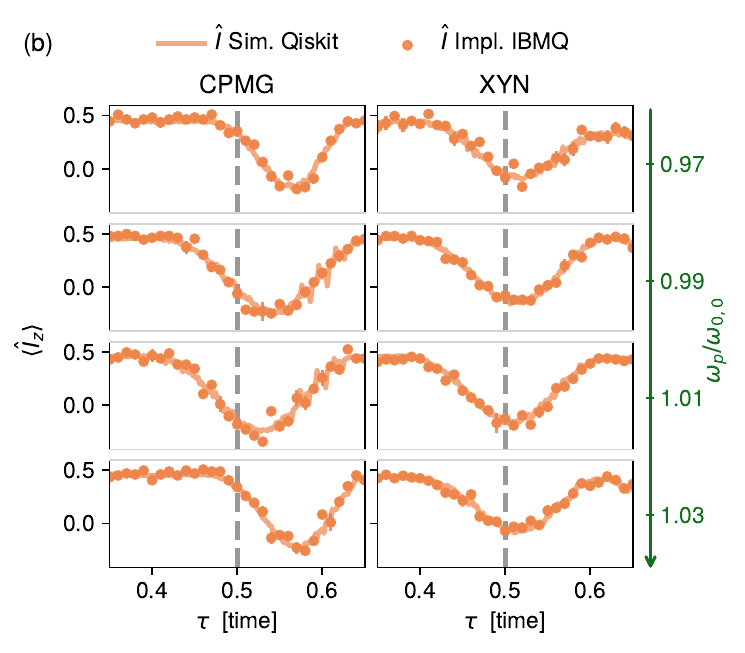}
	\caption[DD Resonances Under Pulse Errors with IBMQ]{IBMQ simulation and implementation of CPMG-8 and XY8 sequences under \textbf{(a)} pulse length error $t_p/t_\pi$ and \textbf{(b)} pulse frequency error $\omega_p/\omega_{0,0}$.
Both sequences present the same resonance feature, but as the pulse errors become larger, the CPMG resonances shift in $\tau$.
	An effect which is not so prominent in the XYN sequence, evidencing a more robust pulse error resilience.
	Small periodical spikes are also observed in the CPMG, demonstrating the robustness of the laboratory frame model, beyond the adoptions of rotating frames via perturbative approximations~\cite{ambiguous_resonances, beyond_RWA}.
	}
	\label{fig:pulse_errors}
\end{figure*}

Another important factor which arises when dealing with multipulse DD sequences are errors in the control field, as they accumulate with each pulse until impairing the application of the sequence.
Here, we consider two types of pulse errors in the control field (Eq.~\ref{eq:IBMQ_H1}): duration of the pulse with respect to a perfect $\pi$-pulse duration $t_p/t_\pi$, and the excitation frequency of the pulse as compared to the resonance of the central qubit $\omega_p/\omega_{0,0}$.

Although the CPMG sequence is one of the most widely used DD sequences, it is particularly prone to such errors~\cite{phd_muller}.
This happens because the $\pi$-pulse is repeatedly applied over the same $x$ axis, causing errors to accumulate fast.
A natural improvement then to the CPMG sequence is to intercalate the $N$ $\pi$-pulses between the $x$ and $y$ axes~\cite{XY8, DD_NV}, which we denote as the XYN sequence.
Notably, the XY8 sequence is widely used in quantum sensing applications~\cite{spurious, ambiguous_resonances}, being composed of eight intercalated pulses anti-symmetrized: [$\pi_x$, $\pi_y$, $\pi_x$, $\pi_y$, $\pi_y$, $\pi_x$, $\pi_y$, $\pi_x$].
In this work however, we do not anti-symmetrize the pulses to simplify the implementation of the DD-gate.
Thus, an XY8 sequence in this work is composed of [$\pi_x$, $\pi_y$, $\pi_x$, $\pi_y$, $\pi_x$, $\pi_y$, $\pi_x$, $\pi_y$].

To compare the CPMG and XYN sequences under different pulse errors, we measured and simulated the target qubit observable $\langle \hat{I}_z \rangle$ in both sequences with $N=8$ pulses.
The same Hamiltonian as in Eq.~\ref{eq:ibmq_H0_2q} was used with coupling term of $A_{zx}=0.2$ [time].
The effect in the sequences of pulse length errors from $0.85 \, t_p/t_\pi$ to $1.15 \, t_p/t_\pi$ while keeping the correct excitation frequency $\, \omega_p/\omega_{0,0}$ are shown in Fig.~\ref{fig:pulse_errors}~(a).
Fig.~\ref{fig:pulse_errors}~(b) shows the sequences under pulse frequency errors from $0.97 \, \omega_p/\omega_{0,0}$ to $1.03 \, \omega_p/\omega_{0,0}$, while keeping the correct pulse duration $t_p=t_\pi$.
Primarily, at small values of errors, both sequences have the same resonance feature at $\tau=0.5$~[time], as expected~\cite{XY8}.
However, as either errors increase, the position of the resonance shifts and the amplitude decreases with the CPMG sequence.
An effect that is less pronounced in the XYN sequence, indicating a more robust pulse errors resilience.

\begin{figure*}[t!]
	\includegraphics[width=\textwidth]{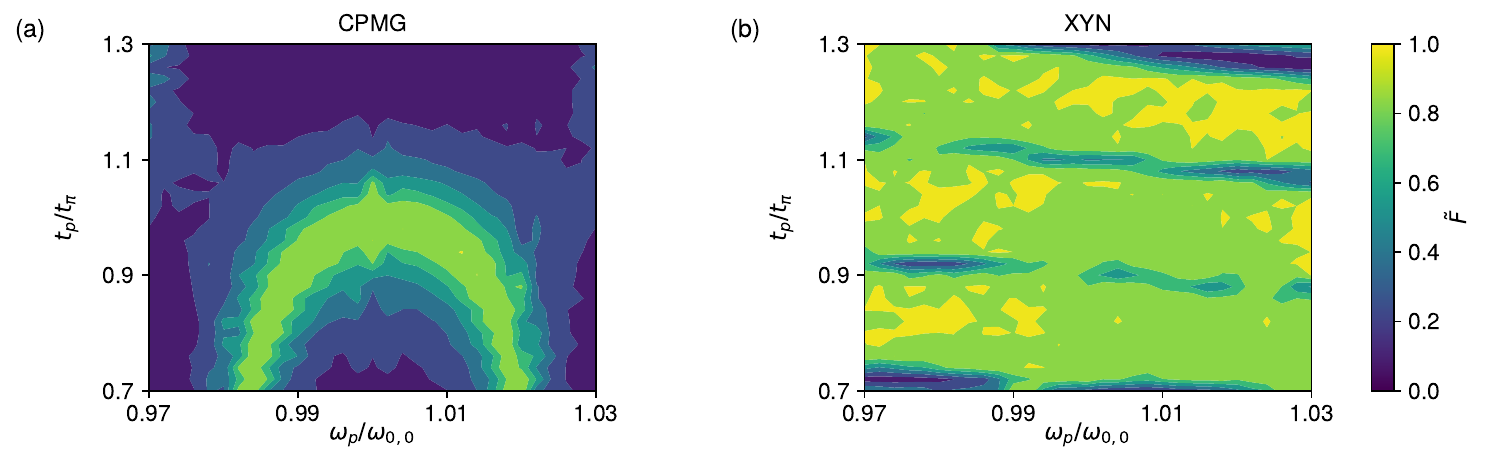}
\caption{
		Simulated pseudo-fidelity $\tilde{F}$ of the DD-gate as a function of pulse errors in IBMQ with \textbf{(a)} CPMG and \textbf{(b)} XYN sequences.
		The CPMG sequence shows a pronounced susceptibility to pulse errors, where small deviations in the control field lead to a substantial decrease of $\tilde{F}$, with a characteristic upside-down `U' shape~\cite{IBMQ_pulse_errors}.
		The XYN sequence, on the other hand, represents a much more robust pulse error resilience, simply by intercalating the $\pi$-pulses in the $x$ and $y$ axes.
		This error resilience could be further improved with anti-symmetrization of the axes~\cite{XY8}, addition of random phases~\cite{RXY8_1, RXY8_2, RXY8_g}, uneven pulse separation~\cite{UDD} or composite pulses~\cite{IBMQ_pulse_errors}.
	}
	\label{fig:pulse_errors_fidelity}
\end{figure*}

The pulse length errors also induce periodical spikes in the resonance peak, similar to as observed in Ref.~\cite{ambiguous_resonances} and when the RWA and adoption of the rotating frame are no longer valid~\cite{beyond_RWA}.
Overall, these results attest to the robustness of this simple model in the laboratory frame, which can describe the behavior of the DD sequences even in the presence of pulse errors.
This incorporation of pulse errors could not be so straightforward or robust under the adoption of rotating frames and RWA.
It could be argued that a shift in the resonance could be easily corrected by adjusting the fixed $\tau$ value for the DD-mediated gate.
However, pulse errors are not generally known prior to the application of the DD sequences.
The assumption of previously knowing the shift to be applied in $\tau$ would thus be a design flaw in the pulse sequence of the DD-gate.

To further characterize the pulse error resilience of the two sequences, we simulate them for several combinations of $t_p/t_\pi$ and $\omega_p/\omega_{0,0}$, varying $N$ and keeping $\tau=0.5$~[time] fixed. 
This results in a series of Rabi-like oscillations as in Fig.~\ref{fig:cpmg_ibmq}~(b), from which we calculate the pseudo-fidelity $\tilde{F}$ of the DD-mediated gate, given by the amplitude of the oscillation (Sec.~\ref{sec:IBMQ_2}).
The results as shown in Fig.~\ref{fig:pulse_errors_fidelity} evidence a massive improvement in pulse error resilience in the XYN as compared to the CPMG.
By simply interleaving the excitation axes in the XYN sequence, pseudo-fidelities above 0.7 can be achieved throughout the considered error values of $0.7 \, t_p/t_\pi$ to $1.3 \, t_p/t_\pi$ and $0.97 \, \omega_p/\omega_{0,0}$ to $1.03 \, \omega_p/\omega_{0,0}$.
Contrastingly, the CPMG sequence suffers from a great reduction in pseudo-fidelity, even with small deviations from the resonant conditions, as observed in other pulse errors resilience studies with IBMQ~\cite{IBMQ_pulse_errors}.

Given this more robust pulse error resilience, the XYN is utilized for the realization of the DD-gate with the NV (Sec.~\ref{sec:NV_2qubit}), where the experimental design is much more susceptible to pulse errors than with IBMQ.
But apart from being more robust, the XYN sequence produces an intrinsically distinct filtering function compared to the CPMG, where the CPMG and XYN sequences have contrasting dynamics depending on the initial state (Sec.~\ref{sec:pol_gen}).
Apart from the anti-symmetrization of the pulses, the pulse error resilience of the XYN sequence for the DD-gate could be further improved with the addition of random phases in each [$\pi_x$, $\pi_y$, $\pi_x$, $\pi_y$] block~\cite{RXY8_1, RXY8_2}, where the random phases can be correlated for even better error suppression~\cite{RXY8_g, ambiguous_resonances}.
Beyond that, asymmetrical pulse separation can be introduced as in Uhrig DD~\cite{UDD}, or composite pulses can be considered~\cite{IBMQ_pulse_errors}.


\section{\textsuperscript{15}NV Hamiltonian Outside the PAS Frame}\label{appendix:NV_PAS}

Although similar, the Hamiltonians of the \textsuperscript{15}NV (Eq.~\ref{eq:H0_NV}) and IBMQ systems under the DD-gates (Eq.~\ref{eq:IBMQ_H0_A}) are not strictly the same.
The difference lies in the fact that the flip-flop terms responsible for the DD-gate with NVs are present within the Zeeman interaction, instead of the hyperfine coupling.
Assuming a small misalignment angle of $\theta_0$ and adopting a reference frame in which $\phi_0=0$, these Zeeman terms of the NV Hamiltonian in the PAS frame are given by
\begin{align*}
	- \gamma^e \mathbf{B}_0 \cdot \hat{\mathbf{S}} &= 
	- \gamma^e B_0 \left( \cos \theta_0 \hat{S}_z + \sin \theta_0 \hat{S}_x \right) \\
	-  \gamma^n \mathbf{B}_0 \cdot \hat{\mathbf{I}} &= 
	- \gamma^n B_0 \left( \cos \theta_0 \hat{I}_z + \sin \theta_0 \hat{I}_x \right).
\end{align*}
If $\theta_0$ and $\mathbf{B}_0$ are too large, such that $|\gamma^e \mathbf{B}_0 \sin \theta_0| \sim D$, these relations are no longer valid, as the basis states for the electronic spin become hybridized~\cite{basis_hybridized}.
To show that these terms are the physical manifestation of the same effect, we assume a new reference frame along the $\mathbf{B}_0$ field, where now the NV axis is misaligned.
In this new frame, the hyperfine coupling tensor is
\begin{equation*}
	\mathbf{A}^\prime = 
	\begin{pmatrix}
		A_{xx}^\prime & 0 & A_{xz}^\prime \\
		0 & A_{yy} & 0 \\
		A_{zx}^\prime & 0 & A_{zz}^\prime
	\end{pmatrix} .
\end{equation*}
With the new elements as~\cite{axis_rot}
\begin{align*}
	A_{xx}^\prime & = A_{xx} \cos^2\theta_0 + A_{zz} \sin^2\theta_0\\
	A_{zz}^\prime &= A_{xx} \sin^2\theta_0 + A_{zz} \cos^2\theta_0 \\
	A_{xz}^\prime = A_{zx}^\prime &=  (A_{xx} - A_{zz}) \sin\theta_0 \cos\theta_0,
\end{align*}
while $A_{yy}$ remains the same.
Clearly then, in this new referential, we have the presence of a $A_{zx}^\prime$ term responsible for DD resonances, as in the model developed for IBMQ.
Although this referential is useful for qualitatively understanding the DD resonances with the $^{15}$NV system, it can be computationally more expensive to simulate than in the PAS frame aligned with the NV axis.
Therefore, we use the PAS frame for the simulations.
\vfil

\begin{figure*}[t!]
	\includegraphics[width=.8\textwidth]{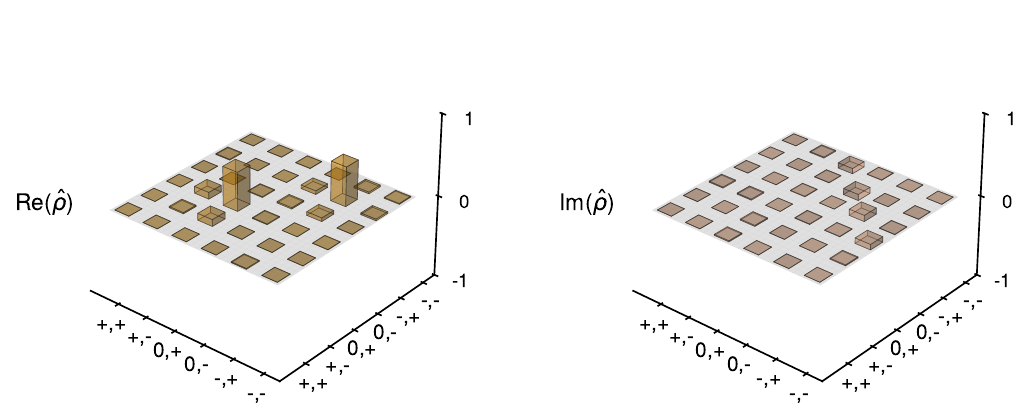}
	\caption[Simulated Quantum State Tomography for Polarization Generation with \textsuperscript{15}NV]{
		Simulated quantum state tomography of the polarization generation DD-gate applied to $^{15}$NV (Sec.~\ref{sec:pol_gen}).
		By applying a CPMG-27 sequence to an initial state $\hat{\rho}_0 = \ket{0}\bra{0}\otimes\hat{\mathds{1}}/2$, the initial polarization of the electronic spin is transferred to the $^{15}$N nuclear spin, with a fidelity of 0.979 in relation to the state $\hat{\mathds{1}}/2 \otimes \ket{m_I=+1/2}\bra{m_I=+1/2}$.
		This method for polarization generation to the target qubit can represent a significant simplification in the experimental design of present techniques~\cite{single_shot_readout, DNP1, DNP2}.
		The states are labeled as $\ket{\pm, \pm} \equiv \ket{m_S = \pm 1} \otimes \ket{m_I = \pm 1/2}$ and $\ket{0, \pm} \equiv \ket{m_S = 0} \otimes \ket{m_I = \pm 1/2}$.
	}
	\label{fig:NV_DM}
\end{figure*}

\pagebreak
%

\end{document}